\documentstyle[12pt,aasms4]{article}
\def\ges{\;_\sim^>\;}

\lefthead{Guerra \& Daly}
\righthead{Central Engines of AGN}

\begin{document}

\title{Central Engines of AGN: Properties of Collimated Outflows
and Applications for Cosmology}

\author{Erick J. Guerra and Ruth A. Daly\altaffilmark{1}}
\affil{Department of Physics, Joseph Henry Laboratories,\\ Princeton
University, Princeton, NJ 08544}

\altaffiltext{1}{National Young Investigator} 

\begin{abstract}

Powerful extended radio galaxies in the 3CR sample
are observed out to redshifts of about 2.  For redshifts
greater than 0.3, the average lobe-lobe size
of these sources decreases monotonically with redshift
for all reasonable cosmological parameter choices.  
This suggests that the characteristic time for which 
an AGN produces highly collimated outflows that
power radio emission is shorter for 
high-redshift sources than it is for low-redshift
sources.
The analysis presented here supports this conclusion.  

The relation between the active lifetime 
and the beam power of powerful extended radio galaxies
is investigated here.  It is found that the data
are accurately described by 
a model in which the active
lifetime of the source, $t_*$, is written as a power-law
in the energy extraction rate, $L_j$. The exponent
of the power law is estimated to be $\beta \simeq 2.1 \pm 0.6$,
where $\beta$ is defined by $t_* \propto L_j^{-\beta/3}$.  
Note that the value of $\beta$ for an Eddington limited
system of zero is excluded by this analysis.  
The fact that $\beta$ is constrained to lie within
a certain range may be used to constrain models of 
large scale jet production and cosmological parameters.     

The comparison of the redshift evolution of characteristic
source sizes with the average lobe-lobe size for powerful
extended 3CR radio galaxies can be used to constrain
cosmological parameters if three empirically estimated 
quantities can be accurately determined for a subset of the 
sources.  As discussed here,
one of these quantities, the lobe propagation velocity, 
is beset by potential biases that
are not completely understood.  The analysis
presented here shows that these biases do not 
significantly affect the 
results on $\beta$, but must be studied in more detail before 
cosmological parameters can be estimated precisely.  
Allowing for the potential biases mentioned above, 
best fits of
the data yield a low of $\Omega_o$, which is about $2\sigma$ away from
a flat, matter-dominated universe.

\end{abstract}


\section{INTRODUCTION}

Active galactic nuclei (AGN) appear in a variety of forms which include
powerful radio emitters produced by highly collimated 
outflows.
The mechanism(s) that produces these
outflows is(are) not known, though several models
have been suggested (e.g. Blandford \& Znajek 1977; 
Blandford \& Payne 1982; Wilson \& Colbert 1995; Lightman
\& Eardley 1974; Rees et al. 1982; Narayan \& Yi 1994, 1995).

Powerful extended radio galaxies can be used to probe 
AGN central engines, as described here and by Daly (1994, 1995).  
Radio 
sources are referred to as 
``powerful'' if their radio powers at 
178 MHz satisfy $P_{178} \ges 3 h^{-2} \times 10^{26}\hbox{ W Hz}^{-1}
\hbox{ sr}^{-1}$, assuming a deceleration parameter $q_o=0$ and
parameterizing Hubble's constant in the usual way:
$H_o=100~h~\hbox{km s}^{-1} \hbox{ kpc}^{-1}$.  These sources are all
on the FRII (edge-brightened) side 
of the FRI-FRII break defined by Fanaroff \& Riley
(1974).  They are
``extended" because their core-lobe separations range from about
$25 ~h^{-1}$ kpc to $240 ~h^{-1}$ kpc (e.g. the radio galaxies 
listed by Wellman, Daly, \& Wan 1997a and 1997b; hereafter WDW97a
and WDW97b), 
which indicate that the outflows are interacting
with intergalactic or intracluster gases.
Only galaxies, and no quasars, 
are considered here to minimize projection effects,
since it is widely believed that powerful extended radio
galaxies have lobes and bridges that lie closer to the plane
of the sky than those of quasars
(see \S 3.1 below, and Wan \& Daly 1997a).

One key to the properties of collimated outflows is the relation 
between the 
rate at which energy is channeled into the outflow, known
as the beam power or luminosity in directed kinetic energy,
and other parameters such as the total time the outflow
exists, and the energy available
to power the outflow.  
In an Eddington limited system, 
the rate of energy extraction, $L_E$, is proportional to the mass of the
central compact object, $M$.  Since the energy available
to power the source
is equal to
the mass of the central compact object times the
emission/accretion efficiency, $E \simeq \eta M$, 
the total time 
the outflow exists is $t_E \simeq E/L_E \propto \eta M/M
\propto \eta$.
Thus, the total time the
outflow exists for an Eddington limited system depends only on $\eta$, and 
not explicitly on $L_E$ or $E$. 

The relation between the beam power and the total time
an AGN produces collimated outflows 
is empirically investigated here for the case of powerful extended radio
galaxies.  
The data can be understood if the total time the
AGN produces collimated outflows 
depends on the beam power, and thus, the outflows 
are not Eddington limited systems.  
The empirical relations obtained here can be used
to constrain models of energy extraction
from AGN, and may provide insight 
on the conditions at the very core of AGN.

The method and model described here can be used to constrain
cosmological parameters if the redshift behavior of 
empirically determined 
quantities can be estimated to relatively high accuracy (Daly 1994).
The constraints on cosmological parameters that can be placed
with the existing data are described in 
\S \ref{sec:cosmo}.  
Modest constraints (i.e.
about 2 $\sigma$) on a flat, matter-dominated
universe may be placed at the present time.  More precise
constaints can be placed when the lobe propagation velocity is 
determined more accurately, or can be observed directly.  
As described in \S \ref{ssec:confit}, uncertainties that affect
the lobe propagation velocity do not
affect, within the errors, 
the estimate of the model parameter $\beta$, which relates
the beam power and the total time
an AGN produces collimated outflows.  

A basic model that describes the observed characteristics 
of powerful extended radio sources rather well is presented in 
\S 2. 
The radio source samples used in the present study, including
their limitations, are described in \S 3.  The application of
the data to the model, and constraints on the relation between
the beam power and the time for which the outflow occurs are 
detailed in \S 4.  The use of current data
to estimate and constrain cosmological parameters
is discussed in \S 5.  A general discussion follows in 
\S 6.  

\section{\label{sec:mod}A MODEL OF THE LOBE SEPARATION IN
POWERFUL EXTENDED RADIO SOURCES}

\subsection{\label{ssec:other}Foundations of the Model}
The foundation of a model which describes the lobe separation of
powerful extended radio sources is discussed in detail
by Daly (1994, 1995).  Powerful extended radio sources 
have
regular, straight bridges and lobes that are thought to propagate
supersonically outward into the ambient gas surrounding them
(\cite{lw84}, hereafter LW84; \cite{al87}, hereafter AL87).
The lobe propagation velocities are computed on the basis of spectral 
aging along the bridges 
(e.g. Myers \& Spangler 1985; Leahy, Muxlow, \& Stephens 1989,
hereafter LMS89;
WDW97a,b).

It should be noted that results from standard spectral aging 
have been criticized recently (Katz-Stone, Rudnick, \& 
Anderson 1993; Rudnick, Katz-Stone, \& Anderson 1994; Rudnick \& 
Katz-Stone 1996; Eilek \& Arendt 1996; Eilek 1996).  These authors 
have argued that syncrhotron aging may not be the correct or 
only explanation for the spectral curvature observed in
the bridges of extended radio sources.  Until more conclusive data
are obtained to settle this controversy, we adopt the practice
of using lobe velocities from spectral aging analysis
since WDW97b find that these velocities and the inferred 
surrounding gas temperatures agree well with results from lobe 
assymetry analysis and X-ray observations.

WDW97a find the data agree well with a 
model in which the bridges of these sources 
expand laterally and the plasma inside undergoes adiabatic expansion.
Their analysis suggests that the backflow 
velocity of the plasma within the bridge is small compared to the lobe advance 
velocity and can be neglected.  It should be noted 
that backflow can be negligible 
along quiescent bridges, but significant near the hotspot,
in agreement with numerical simulations (e.g. Norman 1996).  

The ambient gas density around the lobes of powerful extended radio sources
can be estimated on the basis of ram pressure confinement since the lobes
propagate supersonically (see for example Daly 1994, 1995; WDW97b).
It can be shown
that the number density of the ambient gas, $n_a$, is given by 
\begin{equation}
n_a \propto f(b) {{B_{min}^2}\over{v_L^{2}}}~,
\label{eq:dens}
\end{equation}
where $v_L$ is the lobe propagation velocity, $B_{min}$ 
is the minimum energy magnetic field, $b$ is the offset of the lobe magnetic 
field, $B_L$, from minimum energy magnetic field ($B_L=b B_{min}$), 
and $f(b)=(4/3)b^{-3/2}+b^2$.
WDW97a compute values of $b$ for Cygnus A by comparing 
their estimates of pressure in the lobes, pressure in
the bridge, and mach number of lobe advance to
estimates based on X-ray observations
of gas density, thermal pressure and temperature of 
the surrounding ambient gas.
In particular, each of these three independent estimates
gives $b\simeq 0.25$ for Cygnus A 
(see \S \ref{ssec:datanal}), in agreement with the
result of Carilli et al. (1991).  WDW97b show that the source to
source dispersion in $b$ must be rather small, less than about
$15 \%$.  

Daly (1990) derives an expression for the beam power or the luminosity
in directed kinetic energy, $L_j$, by equating the work done by the lobe
while it propagates into the ambient gas
to the energy supplied by the collimated outflow:
\begin{equation}
L_j\propto n_a a_L^2 v_L^3\propto(v_L/k)^3, \quad\mbox{ where }
k\equiv (n_a a_L^2)^{-1/3}
\label{eq:beam}
\end{equation}
and $a_L$ is the lobe width.
It should be noted that the typical $L_j$ for radio
galaxies in the sample examined below (\S \ref{ssec:dstsamp})
is about
$10^{45}\hbox{ ergs s}^{-1}$ assuming $b=0.25$, and 
it appears that $L_j$ is time
independent for a given source
(Wan, Daly, \& Wellman 1996; Wan \& Daly 1997c).  For 
a roughly constant
$L_j$, the energy supplied by the central engine to power the collimated
outflows is $E_i\sim L_j~t_*$ where $t_*$ is the lifetime of the
collimated outflows.  For $t_*$ between $10^7$ to $10^8$,
which agrees with spectral aging
(e.g. AL87; 
LMS89; Liu, Pooley, and Riley 1992, hereafter LPR92), 
the energy supplied by the central engine is 
$E_i\simeq 10^{60}\hbox{ ergs}\simeq10^6 M_{\odot} c^2$.  For
emission/accretion efficiencies of 0.01 to 0.1, the 
mass of the central compact object would be $10^7 M_{\odot}$ to 
$10^8 M_{\odot}$ which is similar to those discussed
in models of jet formation in powerful FRIIs (e.g. Wilson \&
Colbert 1995).

\subsection{\label{ssec:tvsL}Relation Between Active Lifetime and Beam Power}

The data clearly show that 
the average lobe-lobe size of powerful extended 3CR 
(Bennett 1962)
radio galaxies 
decreases with redshift for $z>0.3$, while the lobe propagation
velocity computed on the basis from spectral aging
tends to increase with redshift and radio power 
(see Figure \ref{fig:phys} in this paper and Figure 19 in 
WDW97a).  Also, it appears that there is no relation
between radio power and lobe-lobe size 
at fixed redshift (e.g. Lacy et al. 1993; Nesser et
al. 1995; Wan \& Daly 1997b).
Since the lobe-lobe size is proportional to
the lobe propagation
velocity times the lifetime, 
it can be inferred that high redshift 3CR galaxies 
have shorter lifetimes than low redshift sources.  The 
higher redshift sources also have 
larger beam powers than the lower redshift sources
since radio power and beam power are roughly proportional 
(Wan, Daly, \& Wellman 1996; Wan \& Daly 1997b).  
A shorter 
lifetime for a more powerful source
is contrary to what is expected in an Eddington
limited outflow, since the Eddington lifetime depends only upon
certain efficiency factors and is independent of 
the beam power and the total energy
of the central compact object.
Though it should be noted that
such an outflow could still be Eddington limited if the efficiency 
factors vary with the beam power.

Let the total time that an
AGN produces two highly collimated oppositely directed 
outflows be $t_*$, and
let the rate of growth of the radio bridges,
$v_L$, be roughly constant over
the source lifetime, which is supported by
spectral aging analyses
(AL87, LPR92, Daly 1994, 1995).  The average or characteristic
lobe-lobe size a source would have if it could be observed
over its entire lifetime is $D_* \simeq v_L~t_*$.  
Following Daly (1994), $t_*$ is related to
the energy extraction rate or beam power, $L_j$, by a power law:
\begin{equation} 
t_* \propto {L_j}^{-\beta/3}~~.  
\label{eq:tofLj}
\end{equation}
This implies that the characteristic source size is
\begin{equation}
D_* \propto k~L_j^{(1-\beta)/3}~\propto 
k^{\beta}~v_L^{1-\beta} ~~,
\label{eq:DofkvL}
\end{equation}
where $k=(n_a a_L^2)^{-1/3}$ (see eq. \ref{eq:beam}).
Clearly, for $\beta
=1$, the average source size depends only upon 
$n_a$ and $a_L$, and is independent of $L_j$,
whereas for $\beta > 1$, the average source size
decreases as the beam power and radio power increase.

The implications of eq. (\ref{eq:tofLj}) are far-reaching. 
It implies that the total energy available to power the
outflow is fixed at some initial value, since
$E_i \simeq L_j~t_* \propto L_j^{(1-\beta/3)}$, assuming that 
$L_j$ is roughly constant over the source lifetime
which is supported by the current data (see \S 2.1).
Note that for $\beta \simeq 3$, the energy extraction
rate $L_j$ is independent of the energy $E_i$ available to 
power the outflow, while for $\beta \simeq 0$, 
the beam power and AGN lifetime are independent and 
$L_j \propto E_i$, analogous to 
an Eddington luminosity.  

In practice, $D_*$ is estimated using the expression introduced by 
Daly (1994), originally referred to as $l_*$:
\begin{equation}
D_* \propto \biggl({1\over{B_L a_L}}\biggr)^{2\beta/3}
{v_L}^{1-\beta/3}~.
\label{eq:Dofobs}
\end{equation}
The normalization is chosen so that $D_*$ at $z\simeq 0$ matches the observed
average lobe-lobe size for sources at this redshift that
are subject to the power cut
discussed in \S \ref{ssec:avsamp}.
The magnetic field strength, $B_L$, in a synchrotron-emitting radio source
and the lobe propagation velocity based on spectral aging (Myers
\& Spangler 1985), $v_L$,
can be estimated using the field strength that
minimizes the total energy in relativistic electrons and magnetic field
(e.g. Miley 1980).
The estimates of $B_L$ and $v_L$ can also be computed using magnetic fields
systematically offset from the minimum energy field,
which is discussed in detail by WDW97a and WDW97b (see 
\S \ref{ssec:datanal}).

Assuming the energy density of the field in the 
radio bridge is large compared to that in the microwave 
background radiation (i.e. synchrotron losses dominate over 
inverse Compton losses) and a spectral index of $\alpha=1$
($S_{\nu}\propto\nu^{-\alpha}$),
it can be shown that 
\begin{equation}
D_* \propto (a_or)^{(4/7-2\beta/3)}~(1+z)^{(23/14-5\beta/6)}~,
\label{eq:Dofcos}
\end{equation}
where $z$ is the redshift of the radio source, and $(a_or)$ is the
coordinate distance to the source. 
The value of $\beta$ indicated by the data for all parameter choices
is $\beta \sim 2$ (see \S \ref{ssec:confit}), and for this value,
$D_* \propto (a_or)^{-0.8}$.
The ratio $<D>/D_*$, where
$<D>$ is the average lobe-lobe size, has a rather strong dependence on 
cosmological parameters via $(a_or)$:  
$<D>/D_* \propto
(a_or)^{1.8}~(1+z)^{-1}$,
for $\beta\sim 2$. The results presented in subsequent sections 
include the effects of inverse Compton scattering with
microwave background photons and the observed spectral indices,
but the assumptions made above show the
dependence of $D_*$ on cosmology in one approximation.

The main hypothesis of the model is that single epoch radio data
can be used to estimate the average lobe-lobe size the source
would have if it could be observed over its entire lifetime,
called $D_*$.  Since very powerful FRII radio 
galaxies form a very homogeneous population, the average size of
a given source should be close to the average size of the
population at the same redshift, as discussed 
in detail in \S \ref{ssec:rat}.
Thus, one test of the model is whether $D_*$ tracks $<D>$
independent of redshift
for a large sample of sources selected using similar 
criteria; this is shown to be the case in \S \ref{ssec:confit}.
If the sources are
sampled randomly during their lifetimes, the 
distribution of $t/t_*$ should be
redshift independent, where $t$ is the age of the
source when it is observed and $t_*$ is the lifetime of
collimated outflows; this is shown to be the case 
in \S \ref{ssec:rat}.  Also, if the 
model correctly describes the data, then the ratio
$<D>/D_*$ should exhibit
much less scatter than the ratio $D/D_*$; this is shown 
to be the case in \S \ref{ssec:rat}.

\section{\label{sec:samp}RADIO GALAXIES EXAMINED} 

\subsection{\label{ssec:avsamp}The Sample Used to Estimate $<D>$}

In order to able to use the model described in \S \ref{ssec:tvsL}, the
sources examined must have lobes that
propagate supersonically, and have no significant backflows or bridge
distortions.  Thus, only FRII radio
galaxies with 
$P_{178}(q_o=0) \ges 3 h^{-2} \times 10^{26} \hbox{ W Hz}^{-1}
\hbox{ sr}^{-1}$ are examined here.
(see \S 1 \& \ref{ssec:other}).
Radio-loud quasars
are not included in this work since the lobe-lobe sizes of radio-loud 
quasars evolve differently with redshift than radio
galaxies (e.g. \cite{s88}), their bridges are more distorted 
than those of radio galaxies (LMS89), and they may suffer from more 
serious projection effects. 

The 3CR radio galaxies with galactic latitude greater than $10^{\circ}$
are completely identified
and all have spectroscopic redshifts.  On the basis of published 178\,MHz
fluxes and spectral indices compiled by Spinrad et al. (1985),
81 radio galaxies from the list compiled by
McCarthy, van Breugel, \& Kapahi (1991) have 
$P_{178}(q_o=0)
\ges 3 h^{-2} \times 10^{26} \hbox{ W Hz}^{-1}\hbox{sr}^{-1}$.
These sources are used here to estimate $<D>$.  Cygnus 
A (3C405), which is not listed by
McCarthy et. al. (1991) due to its low galactic latitude 
($6^{\circ}$), is added to this sample
since it is the closest high
power FRII radio galaxy ($z=0.056$), is well studied (e.g. Carilli et al.
1991), and is included in the $D_*$ sample described below (\S 
\ref{ssec:dstsamp}).  

Table \ref{tb:phys} gives the mean and median lobe-lobe sizes for
the high-power FRII 3CR galaxies assuming a flat, 
matter-dominated universe with ($\Omega_o=1.0,\Omega_{\Lambda}=0$),
an open, curvature-dominated universe with
($\Omega_o=0.1,\Omega_{\Lambda}=0$), and 
a spatially flat universe with a nonzero cosmological constant
($\Omega_o=0.1,\Omega_{\Lambda}=0.9$); here and throughout 
$\Omega_o$ refers to the ratio of the mass
density to the critical value at the current 
epoch, $\Omega_{\Lambda}$ refers to the
ratio of the energy density of a
cosmological constant to the critical density, 
$\Omega_k=1-\Omega_o-\Omega_{\Lambda}$ is the normalized
curvature.
A value of 
$h = 1$ is adopted when a value must be chosen, though $h$ 
scales out of many quantities, and the dependence on $h$ is
typically weak when it does enter.  

Figure \ref{fig:phys} shows the mean 
lobe-lobe sizes from Table 1.
Excluding the lowest redshift bin, an obvious
decrease in the physical size with increasing redshift can be seen
for the cosmological parameters considered.
Note that the lowest redshift, which contains Cygnus A, 
is not an obvious outlier in the comparison of $D_*$ and $<D>$ 
as discussed in \S 4.3 below and by Guerra \& Daly (1996).  

\subsection{\label{ssec:dstsamp}The Sample Used To Estimate $D_*$}

An estimate of $D_*$ requires estimates of the lobe radius, 
the lobe magnetic field strength, and the lobe propagation
velocity of the radio source (see eq. \ref{eq:Dofobs}).  Fourteen 3CR
radio galaxies have enough radio bridge data published to estimate these
parameters, and 
satisfy the power cut described in \S \ref{ssec:other}.
Wellman and Daly (1996), and WDW97a,b 
reanalyzed the radio maps from
LMS89 and LPR92 for their studies of the bridge structure and
gaseous environments of powerful extended radio sources, and computed $v_L$,
$B_L$, and $a_L$ for both data sets in a similar manner.    
Table 
2 lists the 14 radio galaxies from WDW97a,b all of which are used
here.  Note that since they are all high-power 3CR radio galaxies, they are
all included in the larger, full high-power 3CR sample of radio galaxies, 
described in \S \ref{ssec:avsamp}.  
One bridge from 3C239 was excluded by WDW97a,b because its
morphology suggests sideflow (see WDW97a,b for details).
WDW97b compute
values for the lobe half-width, $a_L$, 
measured 10 $h^{-1}$ kpc from the hotspot toward the
host galaxy, the magnetic 
fields measured 10 $h^{-1}$ kpc and 25
$h^{-1}$ kpc from the
hotspot toward the host galaxy, and the lobe propagation
velocity, $v_L$, which is affected by offsets in the magnetic field
from minimum energy conditions and corrections for redshift evolution of
spectral index as discussed below.

\subsubsection{\label{ssec:datanal}
Uncertainties that Affect
Estimates of $D_*$}

The magnetic field in the radio lobe, $B_L$, enters into the
estimate of the lobe pressure, $P_L$, and the magnetic field strength
of the radio bridge, $B_B$, and $B_L$ enter into the estimate of the 
lobe propagation velocity.  Both values affect estimate of
the ambient gas 
density, $n_a$ (WDW97b).
Several observations indicate that the magnetic field strength
in powerful extended radio sources
is less than the minimum energy magnetic field.
Expressing the true magnetic field as $B=bB_{min}$, where 
$B_{min}$ is the minimum energy magnetic field, 
Carilli et al. (1991) find that
$b=0.3$ in order for ram pressure confinement
of the lobes of Cygnus A to be consistent with X-ray measurements of the
ambient gas density.
Perley \& Taylor (1991) find a similar value for 3C295 based
on ram pressure confinement of the lobes, 
while Feigelson et al. (1995) and Kaneda et al. (1995) find
similar values for Fornax A by comparing the radio emission
with X-ray produced by inverse Compton scattering of microwave 
background photons with the relativistic electrons that produce the 
radio emission.  WDW97a use 3 independent and
complementary methods
to estimate $b$ in Cygnus A, and their results agree with the
Carilli et al. (1991) result.
WDW97b also show that the source to source
dispersion in $b$ must be small, less than about 15\%.  
However, both $b=0.25$ and
$b=1.0$ cases are examined below for completeness
(\S \ref{ssec:confit}, Tables 3-5).

Another uncertainty that affects 
$D_*$ is whether the evolution of the radio spectral 
index with redshift introduces a systematic error on 
the lobe propagation velocity.  
It has been noted for some time that the radio spectral index
of the 3CR sample and of other samples increases systematically
with redshift;  this could be due to spectral curvature or due to
other causes 
(e.g. R\"{o}ttgering et al. 1994).  The radio spectral index is
an important ingredient in estimating the lobe propagation
velocity via the effects of spectral
aging of relativistic electrons. Though it should be noted that for 
$\beta\sim2$, which is indicated by the data independent
of systematic effects on the lobe propagation velocity 
(\S \ref{ssec:confit}),
$D_*\propto (a_LB_L)^{-4/3}~v_L^{1/3}$, which is a fairly weak dependence
on $v_L$.  
It is not clear whether the data should
be corrected for the systematic increase of the radio spectral
index with redshift.  
This correction, referred to as the $\alpha$-$z$ correction,
does not change the low-redshift velocities, but
decreases the high-redshift velocities by, at
most, a factor of two (see WDW97a, Figure 19).  
Results obtained both with and without the $\alpha$-$z$
correction are examined below
(\S \ref{ssec:confit}, Tables 3-5).

\section{\label{sec:motest}EXAMINING THE MODEL}

\subsection{\label{ssec:dstar}Computed $D_*$'s}

Each bridge has a characteristic core-lobe
length, $r_*\simeq D_* / 2$, where $D_*$ is obtained using 
eq.(\ref{eq:Dofobs}) and inputing the values of $v_L$, $a_L$, 
and $B_L$ from each bridge.  Note that one can estimate $D_*$ using the 
average of the input parameters in eq. (\ref{eq:Dofobs}) over
both bridges, or
by adding $r_*$ for both bridges. 
Both methods would be equivalent if the sources were completely symmetric, 
but the latter method accounts for any slight asymmetry
in these sources and is used here.
As an example for one set of parameter choices,
Table \ref{tb:samp} lists $r_*$ computed using the results 
for the radio galaxies from WDW97b, and assuming $b=0.25$ without the
$\alpha$-$z$ correction, ($\Omega_o=0.1, 
\Omega_{\Lambda}=0$), and $\beta=2.0$.
This example is chosen because ($\Omega_o=0.1, 
\Omega_{\Lambda}=0$) is the moderate choice of the
three example cosmologies introduced in Table 1, and $\beta=2.0$ is the value
of $\beta$ indicated by the data for all
parameter choices (\S \ref{ssec:confit}).
Estimates of $D_*$ throughout this paper
are computed by adding both $r_*$ values, except for 3C239 which has $r_*$
for only one bridge and has $D_*$ set equal to twice the available $r_*$.  
The normalization of $D_*$
is chosen so that the
ratio of the lobe-lobe 
$D_*$ for Cygnus A (3C405) to $<D>$ for the lowest redshift
bin is unity. 

Figure \ref{fig:dst}a shows 
$D_*$ as a function of $(1+z)$ assuming ($\Omega_o=0.1,\Omega_{\Lambda}=0$),
$b=0.25$ without the $\alpha$-$z$ correction, and
$\beta=0.0$
which is indicative of an Eddington limited system; Figure \ref{fig:dst}b
shows the same except it is assumed that $\beta=2.0$ which
is the value indicated by the data 
for all parameter choices in \S \ref{ssec:confit}.
In these figures and all other figures in this work, 
diamond symbols represent
radio galaxies from the LMS89 sample, and star symbols
represent radio galaxies from the LPR92 sample.
These figures show that $D_*$ estimates for LMS89 and LPR92
sources agree well with each other
for the range of $\beta$ considered, although the 2 samples are
more similar for $\beta=2$ than for $\beta=0$.
Note that the LPR92
sources have smaller angular extent than the LMS89 sources (Table 2), 
and have smaller lobe-lobe sizes which implies they are younger 
(see \S \ref{ssec:rat} below).
The fact that 
the smaller sources from LPR92 and the larger 
sources from LMS89 give
consistent $D_*$ at similar redshifts, indicates that the model 
discussed in \S \ref{ssec:tvsL} is yielding 
a characteristic length that is independent of when the source 
is observed during its lifetime.  

Figure \ref{fig:dst}a shows that
$D_*$ increases as redshift 
increases for $\beta=0.0$, which differs 
from the redshift dependence of the average
lobe-lobe sizes (see Figure \ref{fig:phys}). 
Figure \ref{fig:dst}b shows
that, ignoring the lowest redshift bin (Cygnus A),
$D_*$ decreases with increasing redshift
for $\beta=2.0$.
The general trend is that increasing $\beta$ makes the slope of
$D_*$ vs. $(1+z)$ decrease (i.e. become more negative) excluding
the lowest redshift bin.  Thus, for $\beta=2.0$, the redshift behavior
of $D_*$ tracks that of $<D>$.

For $\beta=0.0$, $D_*$ for 3C427.1 ($z=0.572$) is similar 
to the $D_*$ for Cygnus A ($z=0.056$),
while for $\beta=2.0$, $D_*$ for 3C427.1 
is similar to those for the rest of the $0.3<z<0.6$ sources.
Excluding this source slightly decreases, by about 0.3,
estimates of $\beta$ (see \S \ref{ssec:confit}).
Cygnus A ($z=0.056$) has 
low values for $D_*$ when compared with most sources in the redshift
interval $0.3<z<0.9$.  This is intriguing since $<D>$ 
has a relatively low value in the 
lowest redshift bin and tracks
the redshift behavior of $D_*$ when Cygnus A 
is included (see Figure \ref{fig:phys}).  Excluding Cygnus A does not affect
the estimates of $\beta$, but could slightly affect 
estimates of $\Omega_o$
using this data set (see Guerra \& Daly 1996).

\subsection{\label{ssec:rat}Comparing $D$ to $D_*$} 

The observed lobe-lobe size of a powerful extended radio source is
$D \simeq 2v_L t$ where $t$ is the 
age of the source when observed, and $v_L$ is the lobe propagation velocity
which is assumed to be roughly constant over the lifetime of a source.
An alternate way of expressing the
size for a given source is 
\begin{equation}
D=2\biggl({t\over{t_*}}\biggr) D_*.
\label{eq:DofDst}
\end{equation}
In a given redshift range ($z-\delta z$ to $z+\delta z$) 
of a sample of radio sources,
the distribution of $D_*$ values should be peaked around some 
central value $D_{*z}$ in a given redshift range 
($z-\delta z$ to $z+\delta z$).
Taking the values of $D_*$ in this redshift range to
be equal to the constant $D_{*z}$, the mean lobe-lobe size is
\begin{equation}
\biggl<D\biggr> = 2\;\biggl<\biggl({t\over{t_*}}\biggr) D_* \biggr> \simeq 
2\;\biggl<{t\over{t_*}}\biggr> \: D_{*z}.
\label{eq:avDofDst}
\end{equation}
If we assume for a given sample that the 
distribution of $t/t_*$, the fraction of a source's
lifetime at which it is observed, is a constant over $t/t_*$, then $<D>
\simeq D_{*z}$.  More generally, if the $t/t_*$ distribution of a sample
does not depend on redshift, then $<D> \propto D_{*z}$, and
the ratio $<D>$/$D_{*z}$ 
should be constant and independent of redshift.  It is this premise
that allows an estimate of $\beta$ and cosmological parameters.

Figure \ref{fig:bias}
shows the ratio of the 
physical size $D$ to the characteristic size
$D_*$ for the sample of radio galaxies examined here, assuming
$b=0.25$ and not including the $\alpha$-$z$ correction.
In Figure \ref{fig:bias}, ($\Omega_o=0.1, \Omega_{\Lambda}=0.0$) is assumed
since it is the moderate choice of the three example cosmologies in Table 1,
and $\beta=2.0$ is
assumed since it is the value indicated by the data
for all other parameter choices (see \ref{ssec:confit}).
Since eq. (\ref{eq:DofDst})
gives $t/t_*\propto D/D_*$ for a
given source with constant $v_L$,
the sample shows no significant $t/t_*$ redshift evolution.
Fitting $D/D_*=a(1+z)^n$
we obtain $a = 1.15 \pm 0.07$ and
$n= -0.04 \pm 0.09$; similar results are found for 
$b=1.0$ and
including the $\alpha$-$z$ correction.
Thus, the data are consistent with the
sources being observed at random times during their lifetimes.
The reduced $\chi^2$ of the fit is
41 which is due to
the large scatter of $D/D_*$.  It should be
noted that the scatter in $D/D_*$ is much greater than that of
$<D>/D_*$ (see Figure \ref{fig:ratio}b),
which indicates that though the sources are sampled over a broad range 
of fractional ages, sources at a given redshift 
yield similar $D_*$. 

\subsection{\label{ssec:confit}Constraints Obtained by Comparing $<D>$ to
$D_*$ at a Given Redshift}

A quantitative constraint on the parameter $\beta$ may be
obtained by fitting the ratio
$<D>/D_*$ to a constant, independent of redshift, 
and finding the value of $\beta$ that minimizes
the reduced ${\chi}^2$.  
Table \ref{tb:con} lists
the values of $\beta$ that minimize the reduced $\chi^2$,
$\beta_{min}$, for ($\Omega_o=1.0, \Omega_{\Lambda}=0$),
($\Omega_o=0.1, \Omega_{\Lambda}=0$), and
($\Omega_o=0.1, \Omega_{\Lambda}=0.9$).
Four cases are examined in Table \ref{tb:con}:
(A) $b=0.25$ without the $\alpha$-$z$ correction, (B) $b=0.25$ with the
$\alpha$-$z$ correction, (C) $b=1.0$ without the $\alpha$-$z$
correction, and (D) $b=1.0$ with the $\alpha$-$z$ correction. 
Error estimates for the fitted parameters here and
thoughout are the 68\% confidence
intervals that correspond to 
the regions of
the parameter space within
which the $\chi^2$ increases by no more than 1.0 from the minimum
value (e.g. Press, Teukolsky,
Vettering, \& Flannery 1992; \S 15.6).  Although the errors
that enter into computing the $\chi^2$ may not be Gaussian, we shall
assume Gaussian errors as a rough approximation
in order to use the $\chi^2$ as an
estimate of errors for fitted parameters.
The $\beta_{min}$'s from Table \ref{tb:con}
are all consistent within
$2\sigma$ for
all the cases (A-D) and the three cosmologies examined. 
A range of about 1.5 to 2.6 is found for $\beta_{min}$.

Figures \ref{fig:surfa}a \&\ref{fig:surfa}b 
show contour plots of the reduced $\chi^2$
as a function of $\beta$ and $\Omega_o$, for case A described above, assuming
$\Omega_{\Lambda}=0$ and 
$\Omega_k=0$ respectively.  It is clear from these figures that the
constraints on $\beta$ are not significantly affected by the assumed
cosmology, and a consistent range of $\beta_{min}$ from 1.5 to 2.75 emerges.  
Most significantly, $\beta=0$ gives a reduced $\chi^2$ just slightly greater
than 6, for all cosmologies shown.  Thus, taken at face value, the data 
are not consistent with Eddington limited outflows in 
powerful extended radio sources.  Contour plots
for the cases B, C, and D described above are also shown
in Figures \ref{fig:surfb}a-b, \ref{fig:surfc}a-b, and \ref{fig:surfd}a-b 
respectively.  The constraints on $\beta$ are not significantly different for
the four cases (A-D) examined, and all cases give a high reduced $\chi^2$,
greater than 3,
for $\beta=0$.  Note that these figures only depict the range of $\Omega_o$
from 0 to 1, but fits have been extended outside this range, including 
$\Omega_o<0$ which gives divergent values of the coordinate distance, $(a_or)$,
for $k=0$ (see Table \ref{tb:om}).

Table \ref{tb:om} shows the best fit values of $\Omega_o$ and $\beta$ for 
a constant $<D>$/$D_*$ independent of redshift.
The fits for 
$\Omega_o$ vary by about 0.2 to 0.4 depending on the case
(A-D) described above, which follows from Figures 
\ref{fig:surfa}-\ref{fig:surfd}.  A value of $\Omega_o$
less than unity is indicated by these fits for all parameter choices, 
and $\Omega_o=1$ is inconsistent at about the $2\sigma$ level.  These
constraints on $\Omega_o$ are as strong as those from studies of
large scale velocity fields which indicate that
$\Omega_o>0.3$ at about the $2.4\sigma$ level 
(e.g. Dekel et. al 1993, Dekel \& Rees 1994, Hudson et. al 1995).
The use of radio powerful extended radio galaxies
to constrain cosmological parameters
is discussed further in \S \ref{sec:cosmo} below.

Table \ref{tb:pow} shows the results of 
fitting $<D>$/$D_*$ to a power law in $(1+z)$,
$<D>$/$D_*\propto(1+z)^p$.  A value of $p=0$ is expected for
the correct parameter choices (see \S \ref{ssec:rat}).
Best fit values for $\beta$ and $p$ are shown
for the four cases described above (A-D) and 
three choices of cosmological parameters.  It is interesting
to note that the fits for $\beta$ converge
to similar values for different
cosmologies when a power law fit is allowed.
The fits where 
($\Omega_o=1.0, \Omega_{\Lambda}=0$) are the only ones that give a
significantly non-zero value of $p$, in which case $p$ is about
$2\sigma$ less than zero.
Figures \ref{fig:ratio}a,
\ref{fig:ratio}b, and \ref{fig:ratio}c show the
ratio $<D>$/$D_*$ and the best fit to a constant (the weighted mean in
log-space), assuming $b=0.25$ without the $\alpha$-$z$
correction and $\beta=2.0$, for ($\Omega_o=1.0, \Omega_{\Lambda}=0$),
($\Omega_o=0.1, \Omega_{\Lambda}=0$), and
($\Omega_o=0.1, \Omega_{\Lambda}=0.9$) respectively.
The slopes found in the power law
fits for ($\Omega_o=1.0, \Omega_{\Lambda}=0$)
are apparent in Figure \ref{fig:ratio}a, and
can be explained by the incorrect choice of cosmological parameters.  
This implies that if the universe is flat and matter-dominated,
then there is some redshift evolution that has not been accounted for.

The possibility that certain
sources in the $D_*$ sample may not be representative of 
powerful extended radio sources at their respective redshifts was
considered.
For example, the only low redshift source for which we have
an estimate of $D_*$ is Cygnus A (3C405, $z=0.056$); perhaps Cygnus A is an
unusual source.  A source that is similar to Cygnus A in many respects, 
including size, lobe propagation velocity, surrounding ambient gas density, 
and ambient gas temperature, is 3C427.1 (WDW97a,b).  Also, two
sources, 3C68.2 and 3C239, do not seem to follow the predictions
of adiabatic bridge
expansion, according to WDW97a.  All of the analyses described above have been
repeated while excluding different combinations of these sources
(3C405, 3C427.1, 3C68.2, and 3C239), and the
differences are negligible except for a slight reduction of 
$\beta_{min}$ by about 0.3 when 3C427.1 is excluded holding all
other parameters fixed.  The reduction of
$\beta_{min}$ when 3C427.1 is excluded occurs because lower
$\beta$ tends to make 3C427.1 an outlier (see Figures
\ref{fig:dst}a,b).

The data give an estimate of $\beta$ that is insensitive
to virtually all other parameter choices and data cuts.   
A reasonable estimate of $\beta$ is $\beta=2.1\pm0.6$ based on
Figures \ref{fig:surfa}-\ref{fig:surfd} 
and Tables 3-5.  It is safe to state that
$\beta$ must be between 1 and 3, and that these outflows are not Eddington
limited, $\beta\not= 0$.  The main conclusion 
that can be drawn about cosmological
parameters is
that a flat, matter-dominated universe ($\Omega_o=1.0, \Omega_{\Lambda}=0$)
is about $2\sigma$ 
away from the best fit values (see \S \ref{sec:cosmo} below).

\section{\label{sec:cosmo}CAN POWERFUL EXTENDED RADIO SOURCES BE USED FOR
COSMOLOGY?}

In order to be able to use the model outlined in \S \ref{ssec:tvsL} to estimate
cosmological parameters, accurate estimates of $a_L$, $B_L$, and
$v_L$ are needed (see eq. \ref{eq:Dofobs}). 
Though two inputs to $v_L$
are not completely 
understood at this time, it is still interesting
to apply the method to this data set and consider all possible effects
on $v_L$ to see the implications for cosmological parameters.  

One factor that is not completely understood is 
the offset of magnetic field strength from the
minimum energy value.  
This enters as a scaling factor of $B_L$ and
only affects the redshift behavior of 
$D_*$ through the redshift behavior of $v_L$.  
It turns out that the dependence of $D_*$ on $v_L$ is rather weak
for $\beta \sim 2$ 
(see eq. \ref{eq:Dofobs}):  $D_* \propto
(a_LB_L)^{-4/3}~v_L^{1/3}$, but the effect of the offset from
minimum energy conditions on $v_L$ is considered nonetheless.  
As the magnetic
field strength decreases from the minimum energy value, the
role of inverse Compton cooling begins to become important
relative to the role of synchrotron cooling in these sources.
The affects of both types of cooling on the radio spectral index is used
to estimate spectral ages, and thus $v_L$.
As discussed by WDW97b, for $b = 0.25$, the 2 cooling mechanisms
are of comparable importance for many sources at high redshift.  
This offset
affects both the redshift
behavior of the $v_L$ (shown in Figure 4
of WDW97b), and the dependence of $D_*$ on $(a_o r)$, the coordinate
distance to the source.  

The second effect that may change estimates of the lobe
propagation velocity is the increase of the radio spectral
index with redshift.  It is not clear whether the data should
be corrected to account for the observed
systematic increase of the radio spectral index with redshift
(discussed in detail by WDW97b).  
For 3CR radio 
galaxies, $v_L$ increases with redshift, as do 
the radio power and the radio spectral 
index.  This
increase is well known and has been noted by many authors.  
If, for example, the initial radio spectrum is
not a power law but has some curvature (see
e.g. R\"ottgering et al., 1994),
the 
observed spectral index should be corrected for the observed
redshift evolution as was done for the $\alpha$ 
corrected data (WDW97b).  On the other hand, if the correlation is not due
to, or only partly due to, spectral curvature than such a 
correction is not appropriate.  

To account for these two uncertainties, 
both $b=0.25$ and $b=1$ are considered, 
and both $\alpha$-$z$ corrected and $\alpha$-$z$ uncorrected velocities
are considered (see \S \ref{ssec:confit}).  The 
results are not strongly dependent on these changes
(see Table \ref{tb:om}), as expected since $D_*\propto v_L^{1/3}$.
The effect of using a value of $b=1.0$ instead of $b=0.25$ is to 
increase the best fit $\Omega_o$ by about 0.2
and the effect of including the corrections to
spectral index is to lower the best fit $\Omega_o$ by about 0.2
(see
Table \ref{tb:om}).  A {\it precise} fit for $\Omega_o$ cannot be made until
the parameters that are varied between 
the four cases (A-D) described in \S 4.3 are known.  However,
the current range of $\Omega_o$ allowed can be discussed.

An ideal solution to the uncertainties described above would be to 
identify an independent estimate of the lobe propagation
velocity for all or some of these sources.  
This could be compared with the lobe propagation velocity 
estimated using the spectral aging model used here.  
The velocity can be independently estimated
if either the ambient gas temperature
or the ambient gas density can be estimated.  WDW97a 
have shown that the geometrical shape of the radio bridge
may be used to estimate the Mach number of the lobe, so
the combination of the Mach number and the ambient
gas temperature can be used to estimate the lobe
propagation velocity.  Perhaps observations by AXAF will be used
to measure the ambient gas temperature since its spatial resolution
and spectral coverage could allow it to separate X-rays produced by the AGN from
those produced by thermal bremsstrahlung from the hot ambient gas.
Daly (1994, 1995) and WDW97b 
have shown that the lobe propagation velocity and lobe
pressure may be used to estimate the ambient gas density;
if the ambient gas density can be estimated independently,
the lobe pressure and ambient gas density may be combined
to solve for the lobe propagation velocity.  Studies of the gaseous
environments of these sources using the Sunyaev-Zeldovich effect or 
high resolution AXAF 
measurements may lead to independent estimates of the 
ambient gas density, which may be combined with the lobe pressure
to solve for the lobe propagation velocity.  

\section{\label{sec:disc}DISCUSSION}

A relation between the active lifetime during which an
AGN produces a highly collimated outflows, $t_*$, and the
beam power 
$L_j$, is introduced
which can reconcile the observed evolution of radio source 
size with redshift, and the independence of 
radio source size and radio power at a given redshift 
for powerful extended radio sources (see \S
\ref{ssec:tvsL}).  What emerges from this model
is a characteristic length scale, $D_*$, that can be thought of as the
lobe-lobe size averaged over a source's lifetime, which can be predicted from
measurable quantities such as the lobe width, the lobe propagation
velocity, and the lobe magnetic field strength.  

The model parameter $\beta$, which relates $t_*$
and $L_j$, is estimated 
by comparing the redshift
evolution of $D_*$ to that of the average lobe-lobe sizes for  
powerful extended radio galaxies.
For all the cases and cosmologies
considered, a consistent range of $\beta$ emerges 
(\S \ref{ssec:confit}).
The best fit values allow $\beta$ to range from about 1.5 to 2.75,
so a good rough estimate of $\beta$ is $2.1 \pm 0.6$.  
This is consistent with the range of $\beta=1.5\pm0.5$ 
obtained by Daly (1994, 1995).  

These data are not consistent with $\beta=0$, which is expected
for an Eddington limited system ($t_*$ independent of
$L_j$ as discussed in \S \ref{ssec:tvsL}). In
Figure \ref{fig:dst}a, where $\beta=0.0$ is assumed, the $D_*$ estimates
are clearly increasing with redshift, which is opposite the
evolution of $<D>$ for the high-power 3CR sample described in \S
\ref{ssec:avsamp}.  The data are best fit with 
$\beta>1.0$ which yield reduced
$\chi^2$'s less than 1.5, whereas fits assuming $\beta=0$ yield reduced
$\chi^2$'s greater than 3 (see Figures \ref{fig:surfa}-\ref{fig:surfd}). 
The values of $\beta$ obtained here imply that $D_*$ should decrease
for increasing $L_j$ or $v_L$ (see \S \ref{ssec:tvsL}), which 
is consistent with the 
redshift evolution of the
average sizes of high-power 3CR sources (Figure
\ref{fig:phys}), and the redshift evolution
of $L_j$ and $v_L$ (discussed by Wan \& Daly 1997b).

A flat, matter-dominated universe ($\Omega_o=1.0, \Omega_{\Lambda}=0$)
is allowed at about the $2\sigma$ level for the full range of 
parameter allowed.  
It should be noted that the significance of the constraints 
on $\Omega_o$ obtained
here are similar to those obtained from the study of the 
large scale velocity fields  
(e.g. Dekel et. al 1993, Dekel \& Rees 1994, Hudson et. al 1995).
Stronger cosmological constraints can be placed using this method
when the lobe propagation velocity can be more accurately
determined,  
as discussed in \S \ref{sec:cosmo}, or can be measured directly
by comparing X-ray temperature measurements with the shape of the
radio bridge, as described by WDW97a, or by comparing the X-ray 
density or Sunyaev-Zeldovich pressure with the pressure of the
radio lobe, as described by WDW97b.  

\acknowledgments
Special thanks go to Greg Wellman and Lin Wan for access to their results 
and valuable discussions.  The authors would also
like to thank Mitch Begelman, Roger Blandford, Dave De Young, 
George Djorgovski, 
Stephen Eales, Ed Groth, Paddy Leahy, Simon Lilly, 
Alan Marscher, George Miley, Colin Norman, 
Jerry Ostriker, Lyman Page, Jim Peebles, Rick Perley, Tony
Readhead, Martin Rees, Brigitte Rocca, David Schramm, Rashid
Sunyaev, 
and Dave Wilkinson for helpful
discussions.  
This work was supported in part by the US National Science
Foundation, by a NSF Graduate
Fellowship, the Independent College Fund
of New Jersey, and by a grant from W. M. Wheeler III.  

\clearpage

\begin{deluxetable}{cccccccccccc}
\small
\tablecolumns{11}
\tablewidth{0pc}
\tablecaption{Lobe-Lobe Sizes for $P_{178} \ges 3 h^{-2} \times 10^{26} \hbox{W Hz}^{-1}
\hbox{ sr}^{-1}$ FRII 3CR Galaxies. \label{tb:phys}}
\tablehead{
\colhead{} & \colhead{} & \colhead{} & \colhead{} & \multicolumn{2}{c}{$\Omega_o=1.0,\Omega_{\Lambda}=0.0$} &
\colhead{} & \multicolumn{2}{c}{$\Omega_o=0.1,\Omega_{\Lambda}=0.0$} &
\colhead{} & \multicolumn{2}{c}{$\Omega_o=0.1,\Omega_{\Lambda}=0.9$} \\
\cline{5-6} \cline{8-9}  \cline{11-12} \\
\colhead{} & \colhead{}  & \colhead{no.} & \colhead{} &
\colhead{$<D>$}  & \colhead{$D_{med}$}  &
\colhead{}  &
\colhead{$<D>$} & \colhead{$D_{med}$} &
\colhead{}  &
\colhead{$<D>$} & \colhead{$D_{med}$} \\ 
\colhead{bin} & \colhead{$z$ range} & \colhead{sources} & \colhead{} &
\colhead{(kpc/$h$)} & \colhead{(kpc/$h$)} & 
\colhead{}  &
\colhead{(kpc/$h$)} & \colhead{(kpc/$h$)} & 
\colhead{}  &
\colhead{(kpc/$h$)} & \colhead{(kpc/$h$)} 
}
\startdata
1 & 0.0-0.3 & 3  & & 66$\pm$14  & 52$\pm$14  &  & 68$\pm$14  &
55$\pm$14  & & 72$\pm$13   & 60$\pm$13 \nl
2 & 0.3-0.6 & 15 & & 177$\pm$42 & 160$\pm$43 &  & 197$\pm$47 & 175$\pm$48 &
 & 227$\pm$54 & 200$\pm$55 \nl
3 & 0.6-0.9 & 27 & & 126$\pm$18  & 127$\pm$18  &  & 149$\pm$21  & 147$\pm$21 &  &
179$\pm$25 & 176$\pm$26 \nl 
4 & 0.9-1.2 & 19 & & 91$\pm$21  & 51$\pm$16  &  & 114$\pm$27 & 63$\pm$21 &  &
141$\pm$33 & 79$\pm$26 \nl
5 & 1.2-1.6 & 10 & & 83$\pm$30 & 39$\pm$18  &  & 110$\pm$40 & 51$\pm$25 &  &
137$\pm$50 & 63$\pm$31 \nl
6 & 1.6-2.0 & 8  & & 50$\pm$17  & 37$\pm$18  &  & 71$\pm$24  & 53$\pm$25   
&  & 88$\pm$30  & 66$\pm$31
\enddata
\end{deluxetable}
\clearpage

\begin{deluxetable}{lcccccccc}
\small
\tablewidth{0pc}
\tablecaption{Radio Galaxies with $D_*$ Presented Here.
\label{tb:samp}}
\tablehead{
\colhead{} & \colhead{} & \colhead{} & \colhead{Log[$P_{178}({q_o}=0)$]} 
& \colhead{$\theta$} & \colhead{Map} & \multicolumn{2}{c}{$r_*$}
 & \colhead{$D_*$} \\
\colhead{Source} &  \colhead{z} & bin & 
\colhead{($h^{-2} \hbox{ W Hz}^{-1}\hbox{sr}^{-1}$)} &
\colhead{(arcsec)} & \colhead{Ref.} & \colhead{(kpc/$h$)} &
\colhead{(kpc/$h$)} & \colhead{(kpc/$h$)} }
\startdata
3C405  & 0.056 & 1 & 27.39 & 127 & LMS & 32$\pm$3 & 36$\pm$4 & 68$\pm$5 \nl
3C330  & 0.549 & 2 & 27.01 &  62 & LMS & 82$\pm$13 & 92$\pm$12 & 174$\pm$18\nl
3C427.1 & 0.572 & 2 & 27.08 &  23 & LMS& 49$\pm$5 & 67$\pm$8 & 116$\pm$9 \nl
3C55   & 0.720 & 3 & 27.26 & 71 & LMS & 73$\pm$10 & 101$\pm$17 & 174$\pm$20\nl
3C247  & 0.749 & 3 & 26.89 & 16.7 & LPR & 88$\pm$13 & 97$\pm$12 &185$\pm$17\nl
3C265  & 0.811 & 3 & 27.33 &  78 & LMS & 82$\pm$14 & 100$\pm$17 & 182$\pm$22 \nl
3C289  & 0.967 & 4 & 27.28 &  11.8 & LPR & 58$\pm$6 & 53$\pm$5 & 112$\pm$8 \nl
3C268.1 & 0.974 & 4 & 27.46 & 46 & LMS & 70$\pm$8 & 63$\pm$7 & 133$\pm$10 \nl
3C280  & 0.996 & 4 & 27.60 &  17.7 & LPR & 52$\pm$6 & 64$\pm$6 & 116$\pm$8 \nl
3C356  & 1.079 & 4 & 27.43 &  75 & LMS& 81$\pm$16 & 50$\pm$10 & 131$\pm$18 \nl
3C267  & 1.144 & 4 & 27.58 &  38 & LMS & 54$\pm$9 & 51$\pm$7 & 105$\pm$11\nl
3C68.2  & 1.575 & 5 & 27.85 &  22 & LMS & 111$\pm$29 & 70$\pm$10 
& 181$\pm$31 \nl
3C322  & 1.681 & 6 & 27.84 &  33 & LMS & 37$\pm$4 & 47$\pm$8 & 84$\pm$9 \nl
3C239  & 1.790 & 6 & 28.15 &  13.0 & LPR & 43$\pm$4 & \dag & 87$\pm$6
\enddata
\tablenotetext{\dag}{$r_*$ for only one bridge}
\end{deluxetable}
\clearpage

\begin{deluxetable}{ccccccccc}
\footnotesize
\tablewidth{0pc}
\tablecaption{Best Fit $\beta_{min}$ and $\chi^2_{red}$ Holding
$<D>$/$D_*$ Constant, Independent of $z$.
\label{tb:con}}
\tablehead{
\colhead{}  & \multicolumn{2}{c}{$\Omega_o=1.0,\Omega_{\Lambda}=0.0$} &
\colhead{}  & \multicolumn{2}{c}{$\Omega_o=0.1,\Omega_{\Lambda}=0.0$} &
\colhead{}  & \multicolumn{2}{c}{$\Omega_o=0.1,\Omega_{\Lambda}=0.9$}
\\
\cline{2-3} \cline{5-6} \cline{8-9} \\
\colhead{$case$\tablenotemark{a}} & \colhead{$\beta_{min}$} & \colhead{${\chi^2}_{red}$} &
\colhead{} & \colhead{$\beta_{min}$} & \colhead{${\chi^2}_{red}$}  &
\colhead{} & \colhead{$\beta_{min}$} & \colhead{${\chi^2}_{red}$}
}
\startdata
A & 2.60$\pm$.35 & 1.05 & & 2.30$\pm$.30 & 0.67 & &
2.10$\pm$.30 &
0.69 \nl
B & 2.00$\pm$.35 & 1.84 & & 2.00$\pm$.30 & 1.17 & &
1.95$\pm$.30 &
1.07 \nl
C & 2.35$\pm$.40 & 0.95 & & 2.10$\pm$.35 & 0.67 & &
1.90$\pm$.35 & 0.82
 \nl
D & 1.70$\pm$.30 & 1.37 & & 1.75$\pm$.30 & 1.00 & &
1.75$\pm$.35 &
1.11 \nl
\enddata
\tablenotetext{a}{Refers to the cases: (A) $b=0.25$ without
$\alpha$-$z$ correction, (B) $b=0.25$ with $\alpha$-$z$ correction, (C)
$b=1.0$ without $\alpha$-$z$ correction, (D) $b=1.0$ with $\alpha$-$z$
correction.}
\end{deluxetable}
\clearpage

\begin{deluxetable}{cccccccc}
\footnotesize
\tablewidth{0pc}
\tablecaption{Best Fit $\Omega_o$, $\beta_{min}$, and $\chi^2_{red}$
Fitting $<D>$/$D_*$ to a Constant, Independent of $z$.
\label{tb:om}}
\tablehead{
\colhead{}  & \multicolumn{3}{c}{$\Omega_{\Lambda}=0$} &
\colhead{}  & \multicolumn{3}{c}{$k=0$, $\Omega_{\Lambda}=1-\Omega_o$} \\
\cline{2-4} \cline{6-8} \\
\colhead{case\tablenotemark{a}}  &
\colhead{$\Omega_o$} & \colhead{$\beta_{min}$} &
\colhead{${\chi^2}
_{red}$} &
\colhead{} & \colhead{$\Omega_o$} & \colhead{$\beta_{min}$} &
\colhead{${\chi^2}
_{red}$}
}
\startdata
A & $-0.10^{+.50}_{-.40}$ & 2.15$\pm$.30 & 0.65 & &
$0.20^{+.30}_{-.20}
$ & 2.25$\pm$.30 & 0.67
\nl
B & $-0.35^{+.30}_{-.25}$ & 1.75$\pm$.25 & 0.95 & &
\multicolumn{3}{c}{$\Omega_o<0$} \nl
C & $0.10^{+.50}_{-.35}$ & 2.10$\pm$.35 & 0.67 & &
$0.35^{+.35}_{-.30}
$ & 2.25$\pm$.35 & 0.69
\nl
D & $-0.05^{+.45}_{-.30}$ & 1.75$\pm$.30 & 0.98 & &
$0.20^{+.35}_{-.30}
$ & 1.80$\pm$.35 & 1.07
\enddata
\tablenotetext{a}{Refers to the cases: (A) $b=0.25$ without
$\alpha$-$z$ correction, (B) $b=0.25$ with $\alpha$-$z$ correction, (C)
$b=1.0$ without $\alpha$-$z$ correction, (D) $b=1.0$ with $\alpha$-$z$
correction.}
\end{deluxetable}
\clearpage

\begin{deluxetable}{cccccccccccc}
\tablewidth{0pc}
\scriptsize
\tablecaption{Best Fit $\beta_{min}$, p($\beta_{min}$), and
$\chi^2_{red}$
Setting $<D>$/$D_*=C(1+z)^p$.
\label{tb:pow}}
\tablehead{
\colhead{}  & \multicolumn{3}{c}{$\Omega_o=1.0,\Omega_{\Lambda}=0.0$} &
\colhead{}  & \multicolumn{3}{c}{$\Omega_o=0.1,\Omega_{\Lambda}=0.0$} &
\colhead{}  & \multicolumn{3}{c}{$\Omega_o=0.1,\Omega_{\Lambda}=0.9$}
\\
\cline{2-4} \cline{6-8} \cline{10-12} \\
\colhead{case\tablenotemark{a}} &
\colhead{$\beta_{min}$} & \colhead{p($\beta_{min}$)} &
\colhead{$\chi^2_{red}$} &
\colhead{} & \colhead{$\beta_{min}$} & \colhead{p($\beta_{min}$)}  &
\colhead{$\chi^2_{red}$} &
\colhead{} & \colhead{$\beta_{min}$} & \colhead{p($\beta_{min}$)} &
\colhead{$\chi^2_{red}$}
}
\startdata
A & 2.25$\pm$.35 & -0.75$\pm$.30 & 0.74 &
 & 2.20$\pm$.35 & -0.15$\pm$.30 & 0.71 &
& 2.30$\pm$.35 & 0.30$\pm$.30 & 0.70 \nl
B & 1.85$\pm$.35 & -0.95$\pm$.30 & 1.18 &
 & 1.85$\pm$.35 & -0.40$\pm$.30 & 1.11 &
& 1.90$\pm$.35 & -0.05$\pm$.25 & 1.15 \nl
C & 
2.20$\pm$.40 & -0.60$\pm$.30 & 0.76 &
 & 2.15$\pm$.40 & 0.05$\pm$.30 & 0.72 &
& 2.25$\pm$.40 & 0.50$\pm$.30 & 0.71\nl
D &  1.75$\pm$.35 & -0.60$\pm$.30 & 1.16 &
 & 1.75$\pm$.35 & -0.05$\pm$.30 & 1.08 &
& $1.80\pm$.35 & 0.30$\pm$.30 & 1.11
\enddata
\tablenotetext{a}{Refers to the cases: (A) $b=0.25$ without
$\alpha$-$z$ correction, (B) $b=0.25$ with $\alpha$-$z$ correction, (C)
$b=1.0$ without $\alpha$-$z$ correction, (D) $b=1.0$ with $\alpha$-$z$
correction.}
\end{deluxetable}
\clearpage

\begin{figure}
\figcaption[]{Mean lobe-lobe size $<D>$ vs. $(1+z)$ for 
high-power, $P_{178}(q_o=0) \ges 3 h^{-2} \times 10^{26} 
\hbox{W Hz}^{-1}
\hbox{ sr}^{-1}$, 3CR radio galaxies
for three different cosmologies (see figure legend). \label{fig:phys}}
\end{figure}

\begin{figure}
\figcaption[]{The characteristic length $D_*$ vs. $(1+z)$
computed without the $\alpha$-$z$ correction, and assuming
$\Omega_o=0.1$, $\Omega_{\Lambda}=0$, and $b=0.25$ 
for two choices of $\beta$: 
(a) $\beta=0.0$ and (b) $\beta=2.0$.
Diamonds represent LMS89 sources,
and stars represent LPR92 sources. \label{fig:dst}}
\end{figure}

\begin{figure}
\figcaption[]{
$D/D_*$ 
vs. $(1+z)$, assuming $\Omega_o=0.1$, $\Omega_{\Lambda}=0$,
and $\beta=2.0$, for $b=0.25$ without the $\alpha$-$z$ correction.
Since $v_L$ is 
assumed constant over the lifetime of a 
source, $D/D_* \propto t/t_*$.  Diamonds represent LMS89 sources,
and stars represent LPR92 
sources.  The dashed line is the best fit to a power-law. 
\label{fig:bias}}
\end{figure}

\begin{figure}
\figcaption[]{The {\it reduced $\chi^2$}
as a function of $\beta$ and $\Omega_o$ 
obtained by fitting $<D>/D_*$ to a constant, independent of redshift, 
assuming $b=0.25$ and no $\alpha$-$z$ 
correction: (a) for $\Omega_{\Lambda}=0$
and (b) $\Omega_k=0$.  Integer values of the reduced $\chi^2$ are labeled in
figures; the contours increase in steps of 0.25, with the lowest
contour at a reduced $\chi^2$ of 0.75.  
Note that the reduced $\chi^2 = 1$ contour translates to a $\sigma$ of
about 2 for both figures, assuming Gaussian errors.
\label{fig:surfa}}
\end{figure}

\begin{figure}
\figcaption[]{The {\it reduced $\chi^2$} 
as a function of $\beta$ and $\Omega_o$ 
obtained by fitting $<D>/D_*$ to a constant, independent of redshift, 
assuming $b=0.25$ and with the 
$\alpha$-$z$ correction: (a) for $\Omega_{\Lambda}=0$
and (b) $\Omega_k=0$.  Integer values of the reduced $\chi^2$ are labeled in
figures; the contours increase in steps of 0.25, with the lowest
contour at a reduced $\chi^2$ of 1.25.  
Note that the reduced $\chi^2 = 1.5$ contour translates to a $\sigma$ of
about 2 for both figures, assuming Gaussian errors.
\label{fig:surfb}}
\end{figure}

\begin{figure}
\figcaption[]{The {\it reduced $\chi^2$} 
as a function of $\beta$ and $\Omega_o$ 
obtained by fitting $<D>/D_*$ to a constant, independent of redshift, 
assuming $b=1.0$ and no $\alpha$-$z$ correction: (a) for $\Omega_{\Lambda}=0$
and (b) $\Omega_k=0$.  Integer values of the reduced $\chi^2$ are labeled in
figures; the contours increase in steps of .25, with the lowest
contour at a reduced $\chi^2$ of 0.75.  
Note that the reduced $\chi^2 = 1$ contour translates to a $\sigma$ of
about 2 for both figures, assuming Gaussian errors.
\label{fig:surfc}}
\end{figure}

\begin{figure}
\figcaption[]{The {\it reduced $\chi^2$} 
as a function of $\beta$ and $\Omega_o$ 
obtained by fitting $<D>/D_*$ to a constant, independent of redshift, 
assuming $b=1.0$ and with $\alpha$-$z$ correction: (a) for $\Omega_{\Lambda}=0$
and (b) $\Omega_k=0$.  Integer values of the reduced $\chi^2$ are labeled in
figures;  the contours increase in steps of 0.25, with the lowest 
contour at a reduced $\chi^2$ of 1.25.  
Note that the reduced $\chi^2 = 1.5$ contour translates to a $\sigma$ of
about 2 for both figures, assuming Gaussian errors.
\label{fig:surfd}}
\end{figure}

\begin{figure}
\figcaption[]{
$<D>/D_*$ vs. $(1+z)$, assuming $\beta=2.0$, $b=0.25$, for three 
choices of cosmology:
(a) ($\Omega_o=1.0$, $\Omega_{\Lambda}=0$), (b) ($\Omega_o=0.1$,
$\Omega_{\Lambda}=0$), and (c) ($\Omega_o=0.1$, $\Omega_{\Lambda}=0.9$).
Diamonds represent LMS89 sources,
and stars represent LPR92 sources.  The dashed line is the weighted mean in
log-space of all
$<D>/D_*$ in the figure, assuming $<D>/D_*$ is redshift independent.
\label{fig:ratio}}
\end{figure}

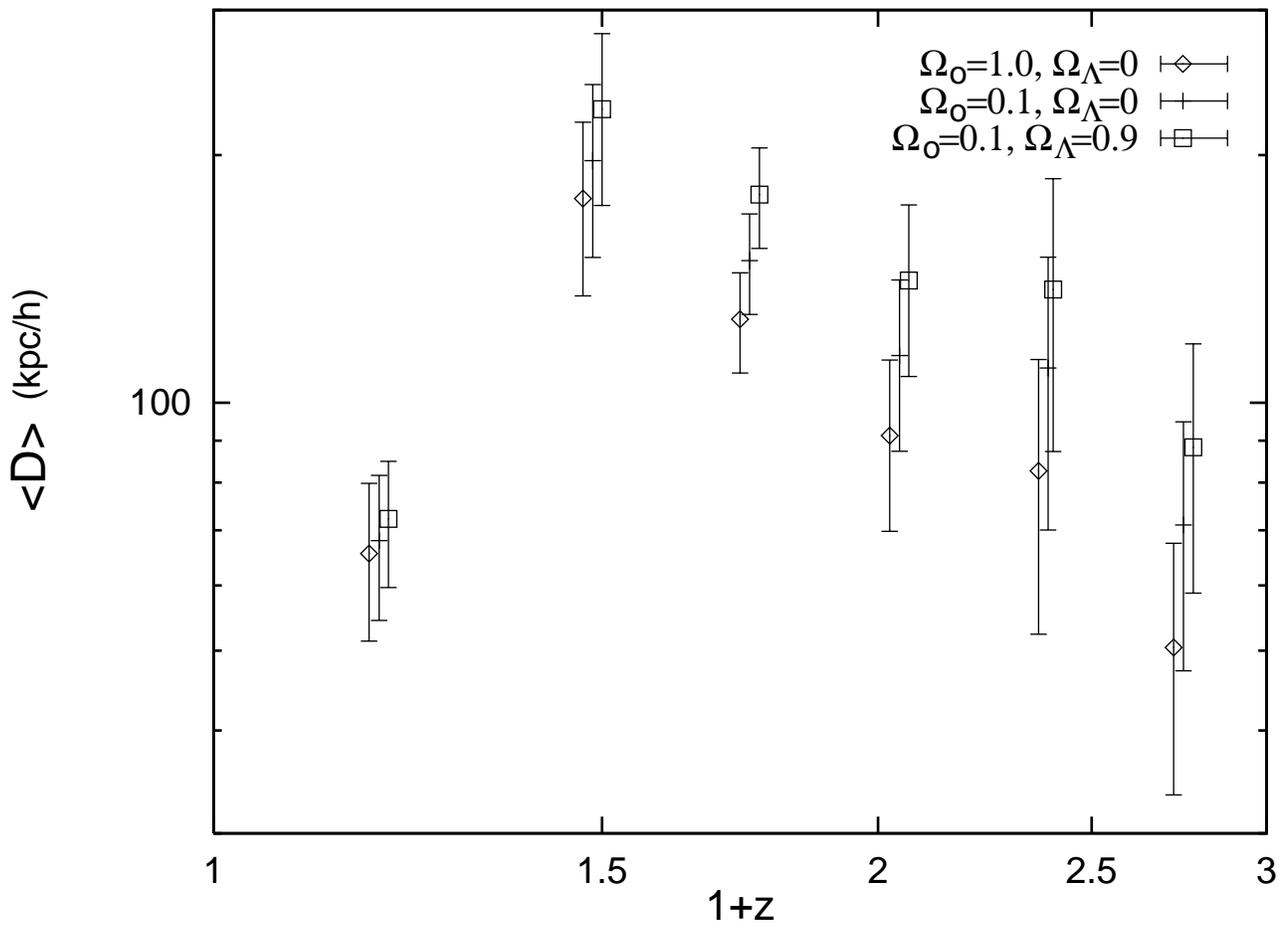
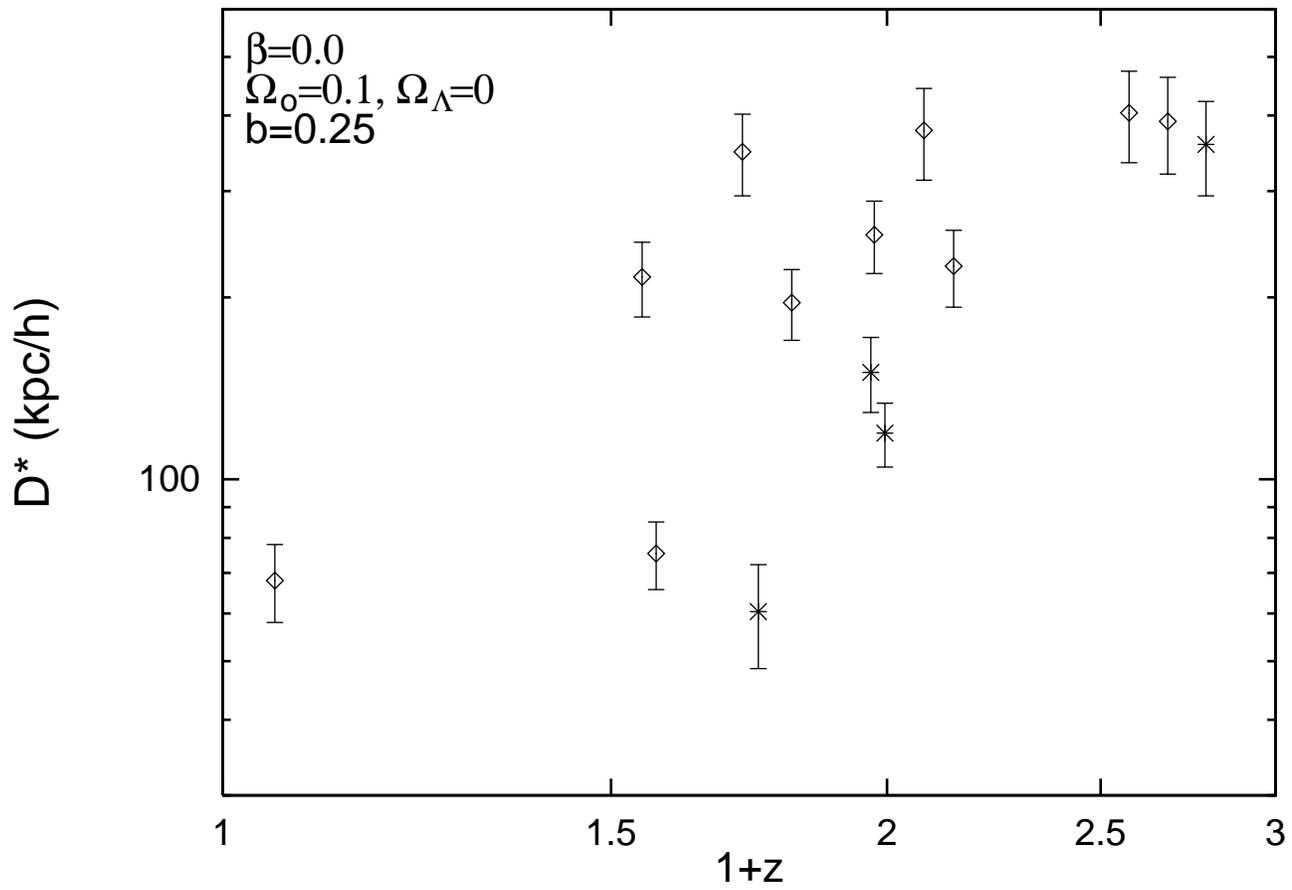
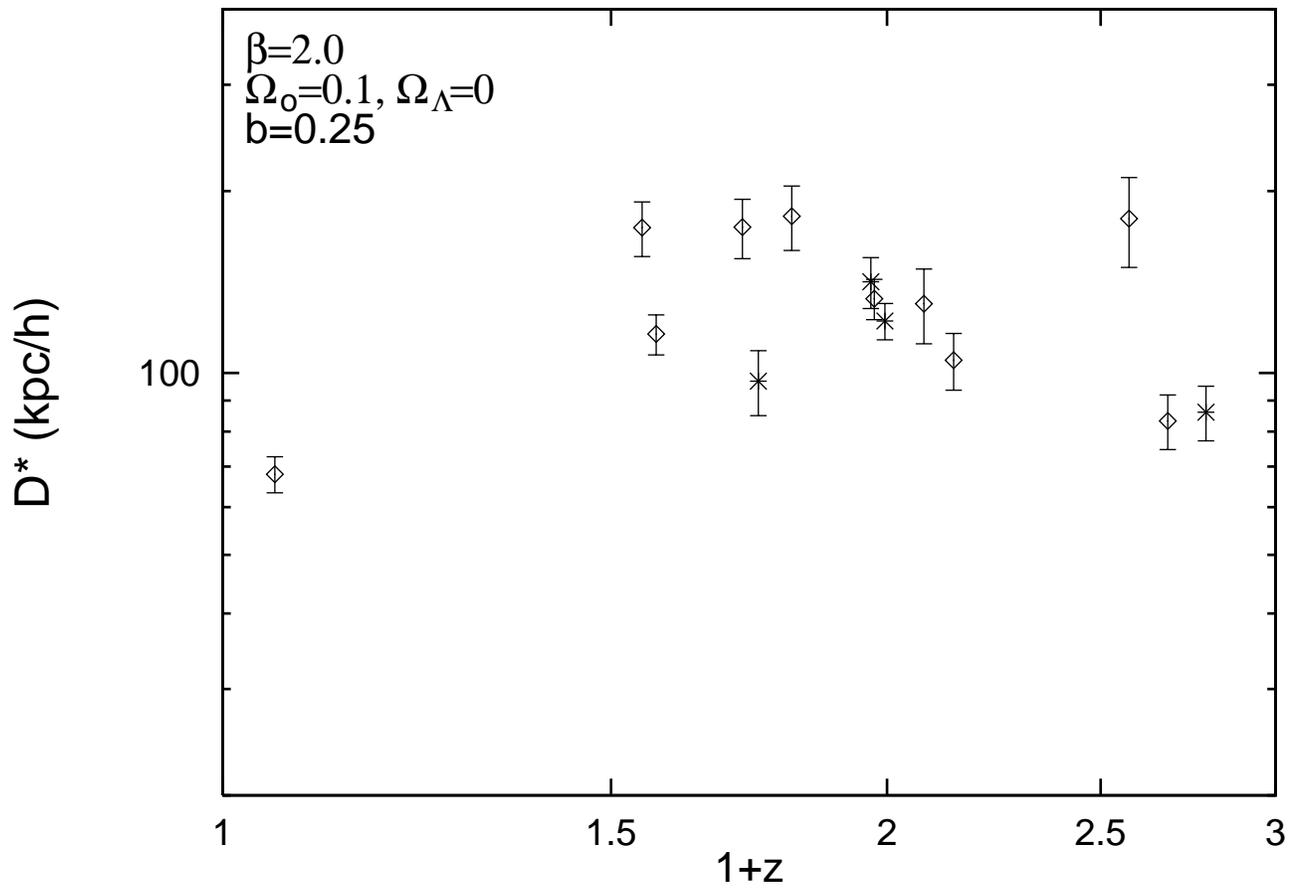
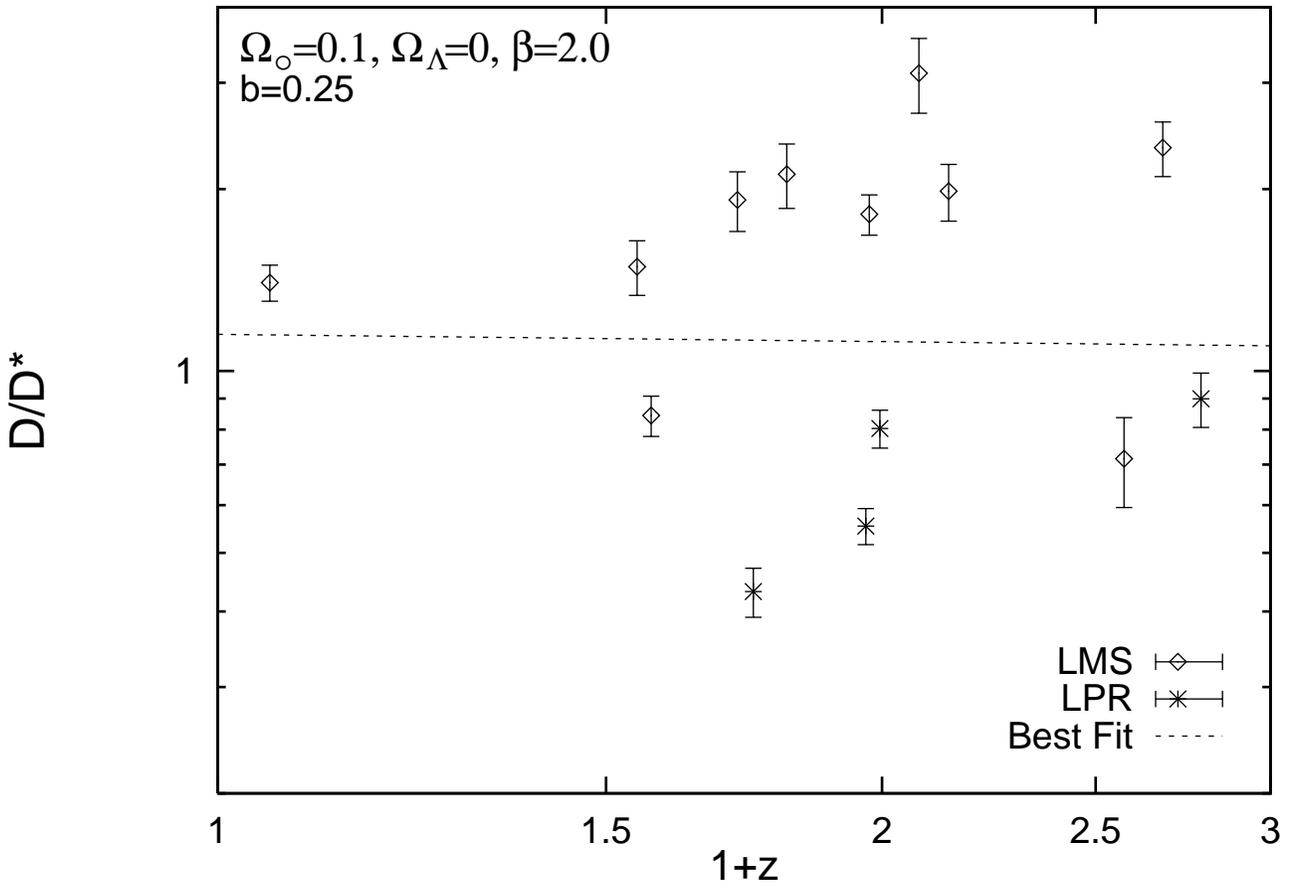
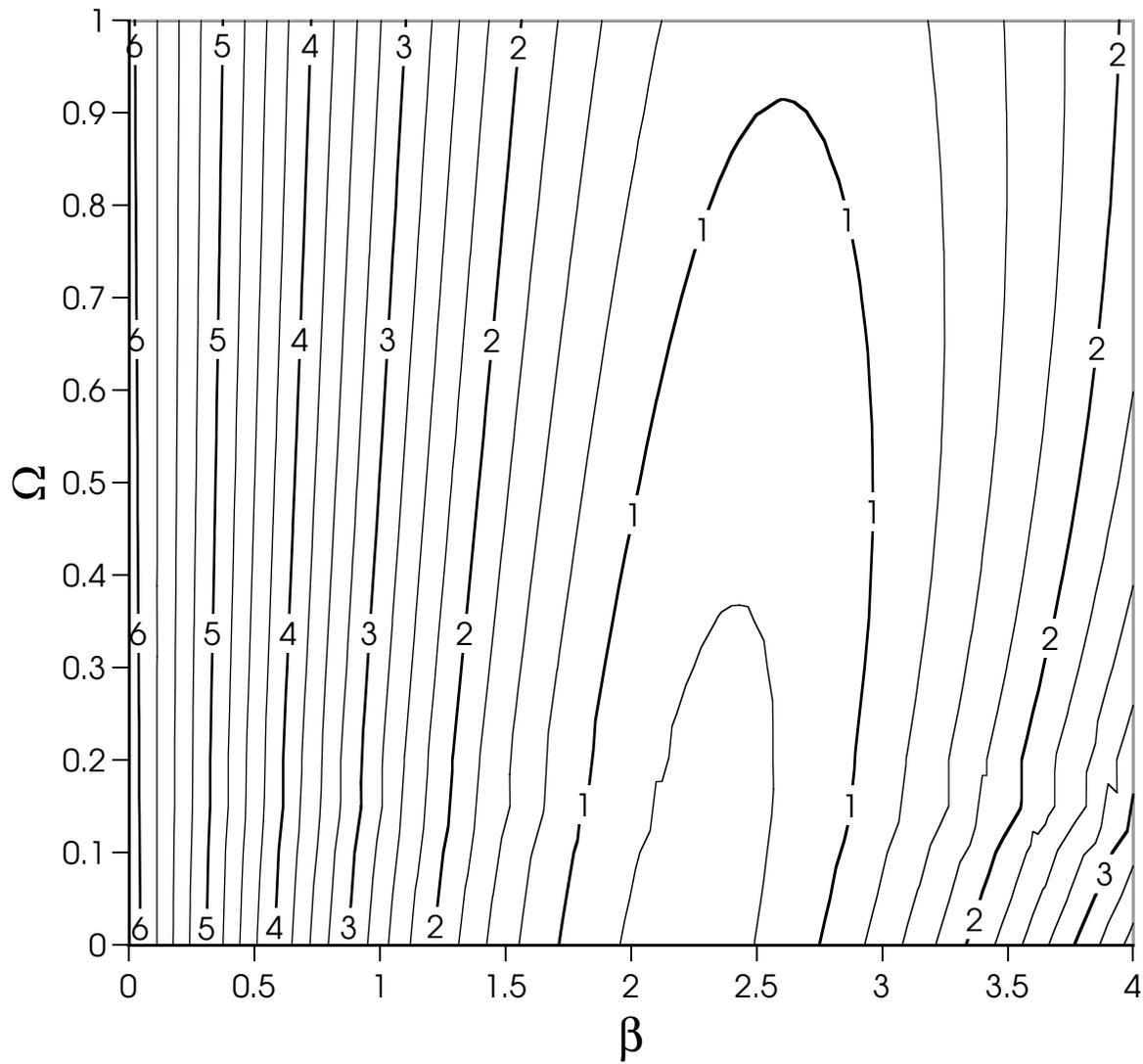
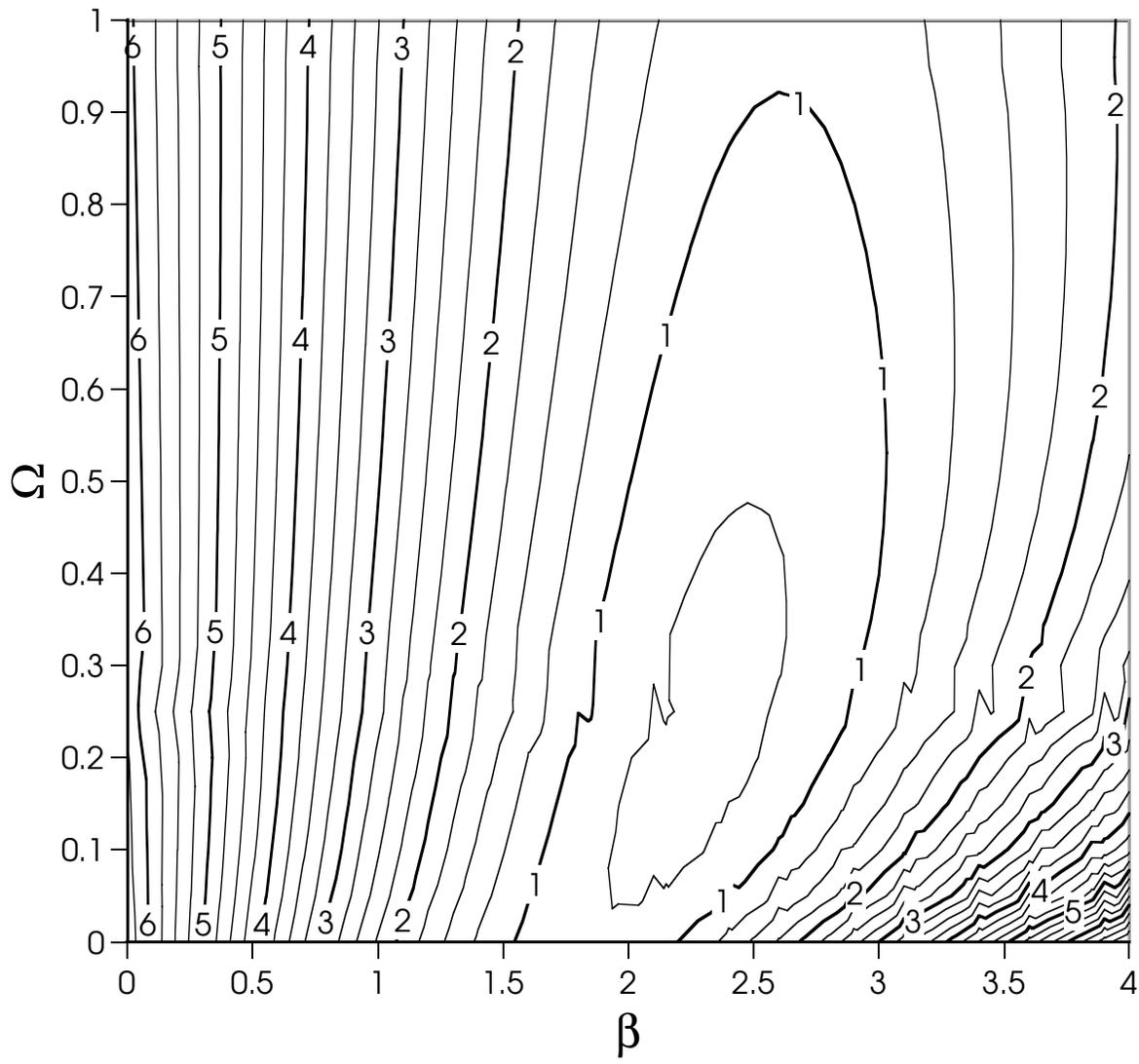
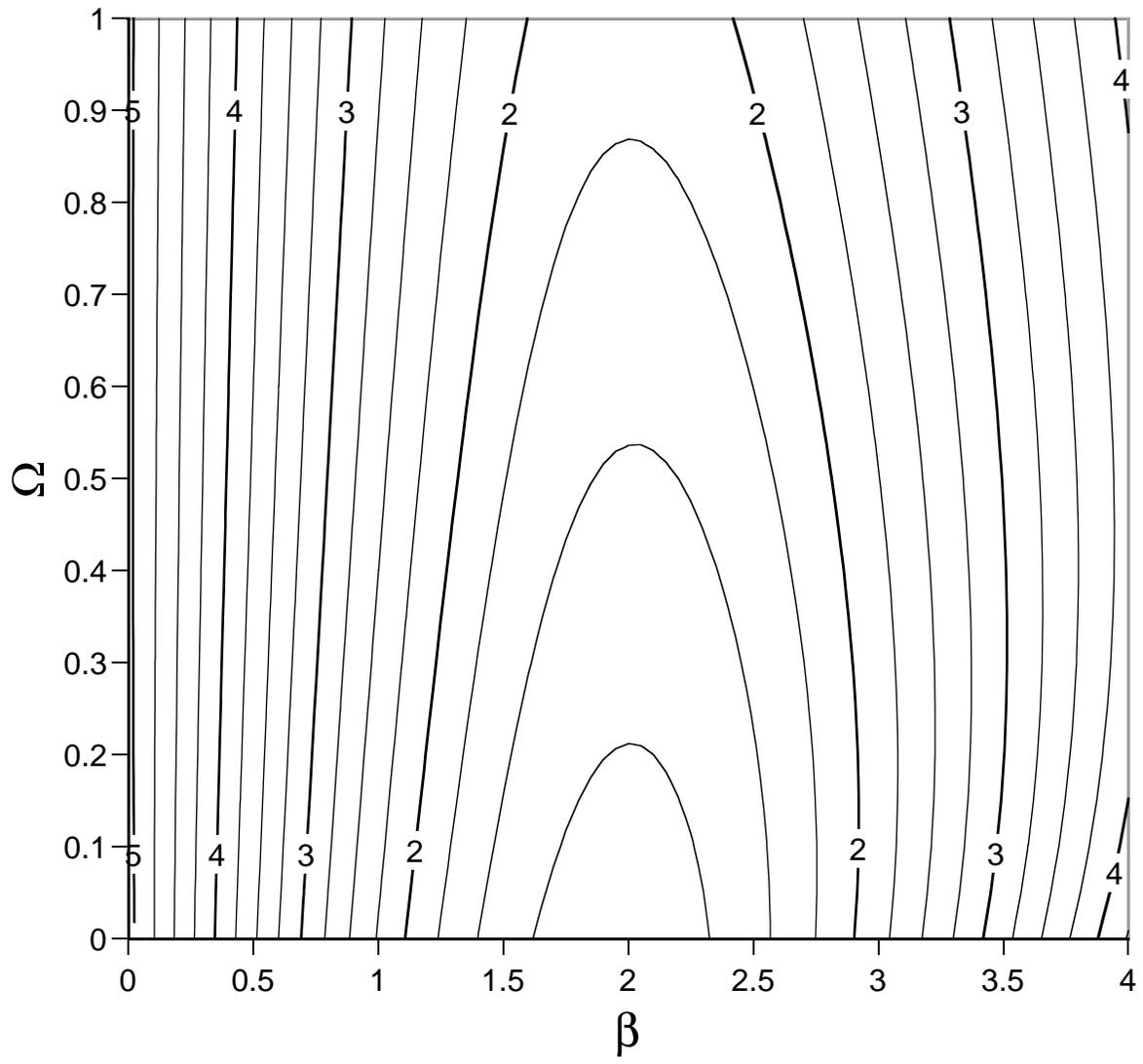
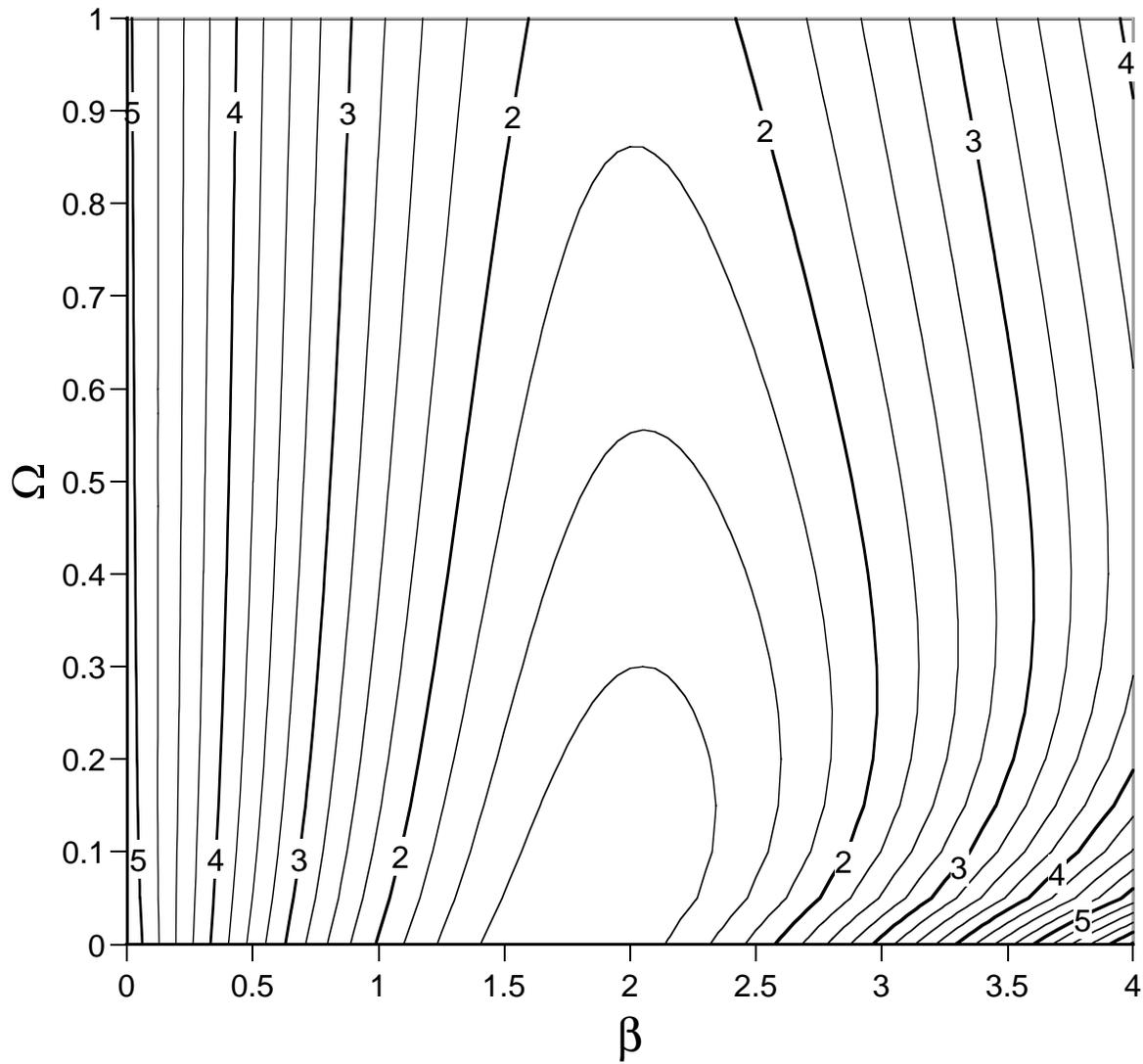
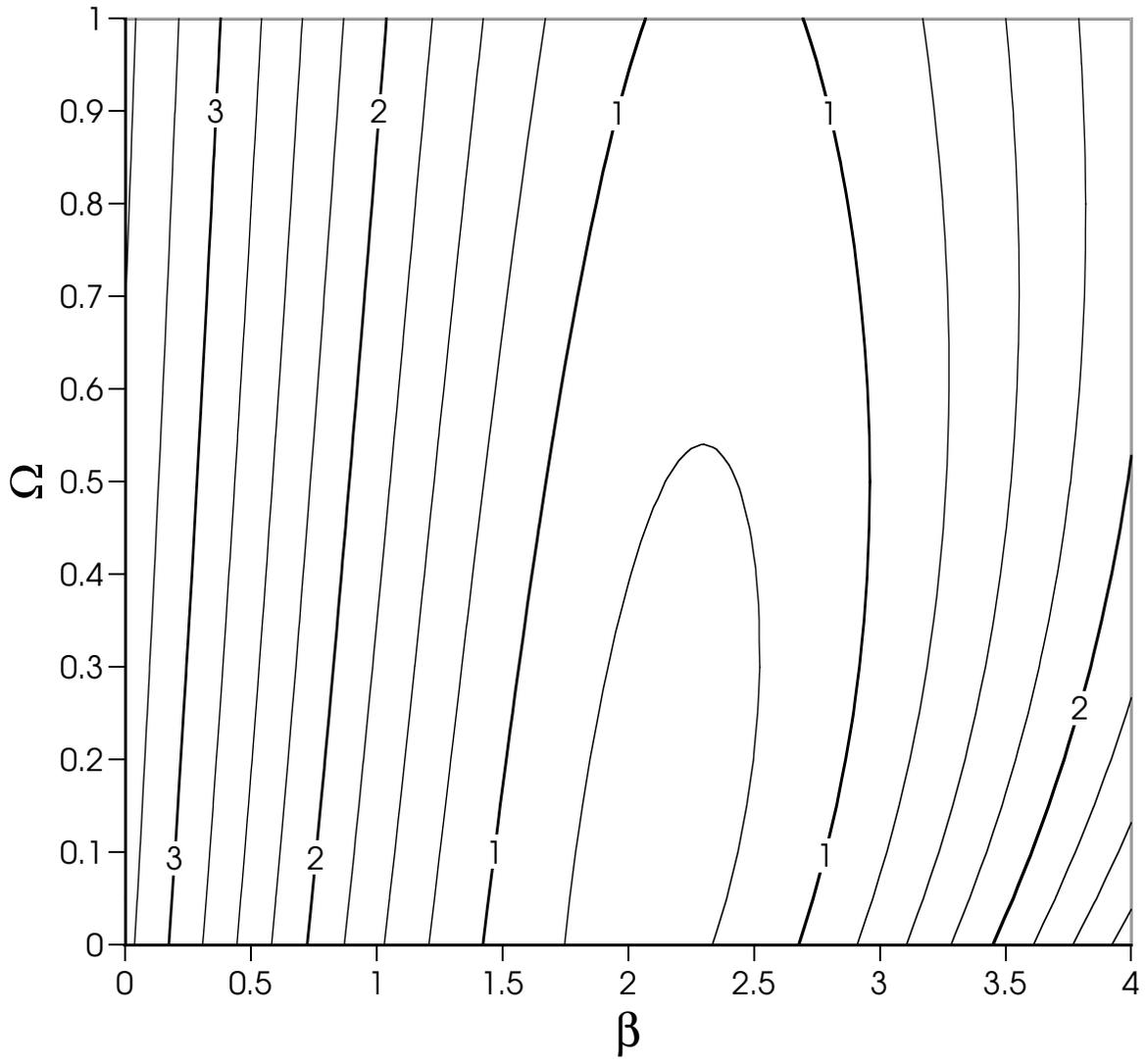
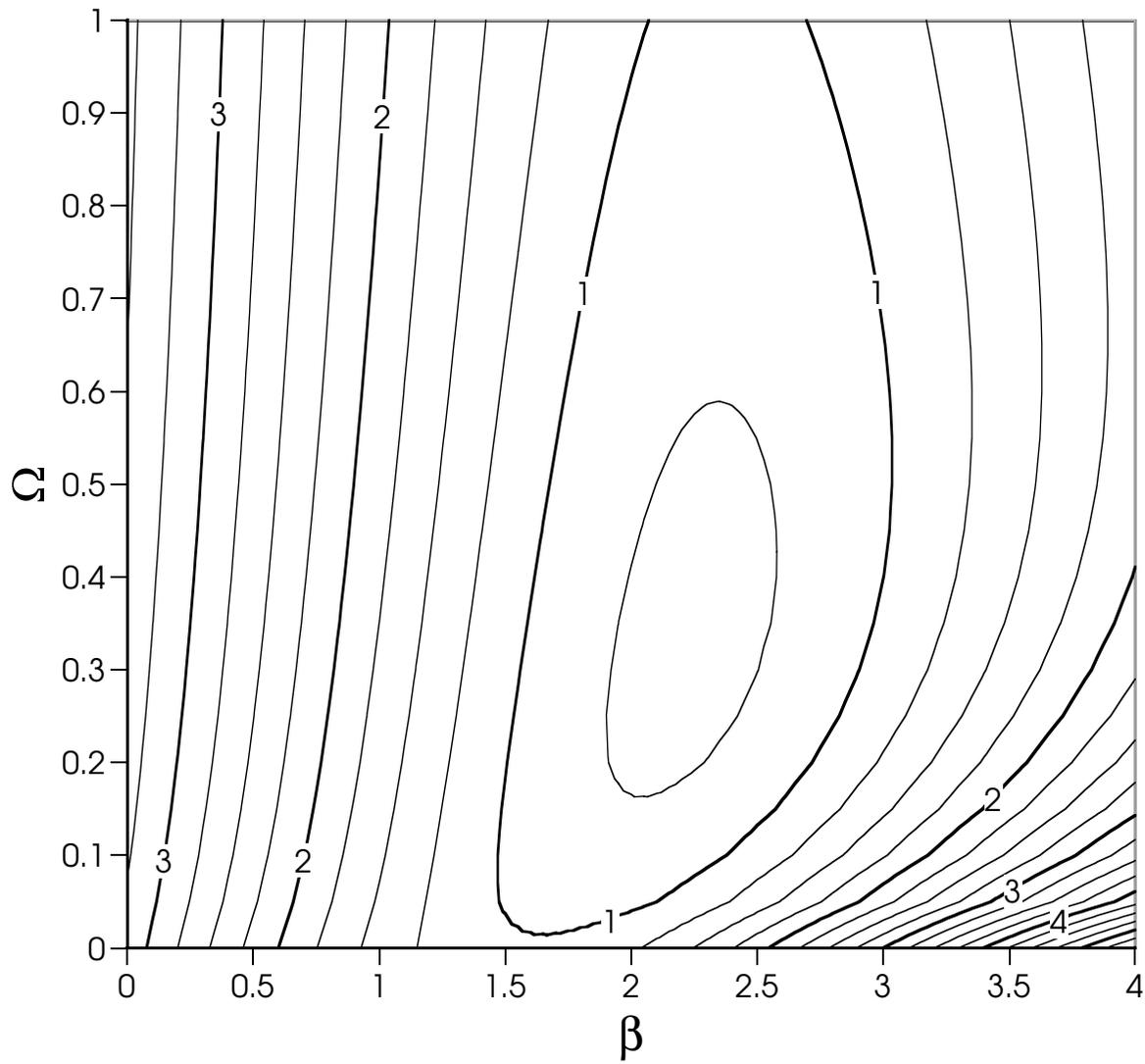
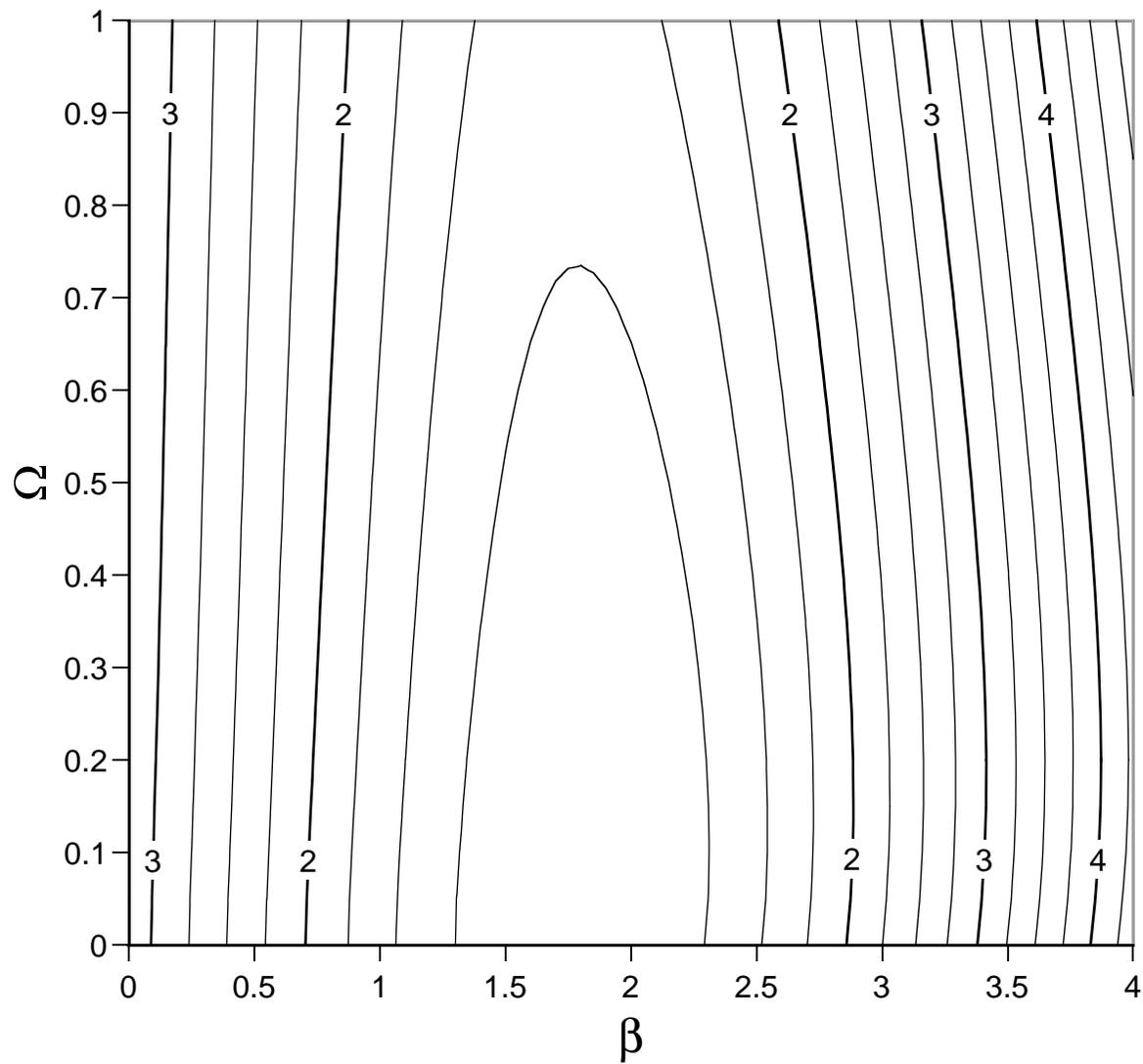
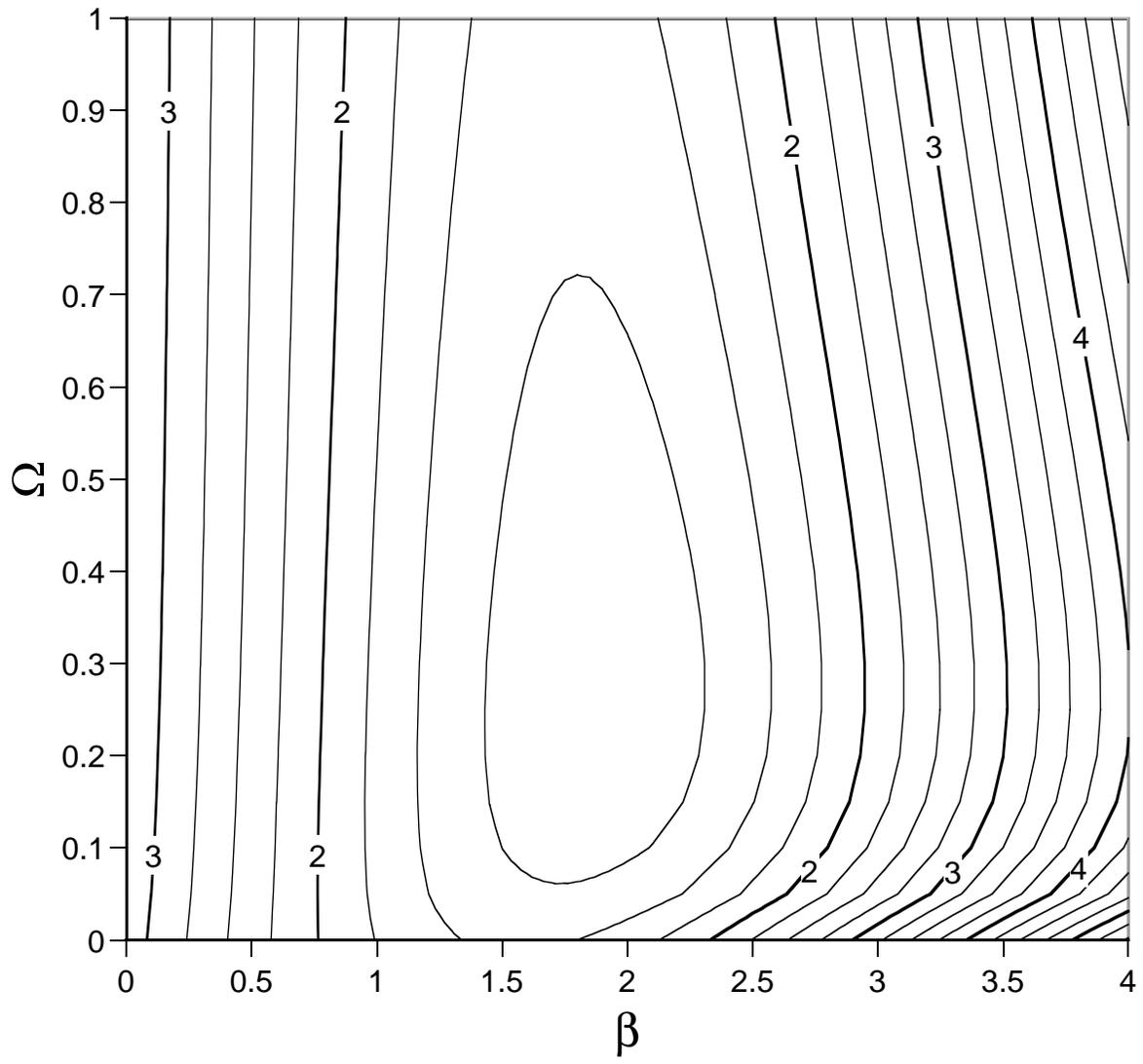
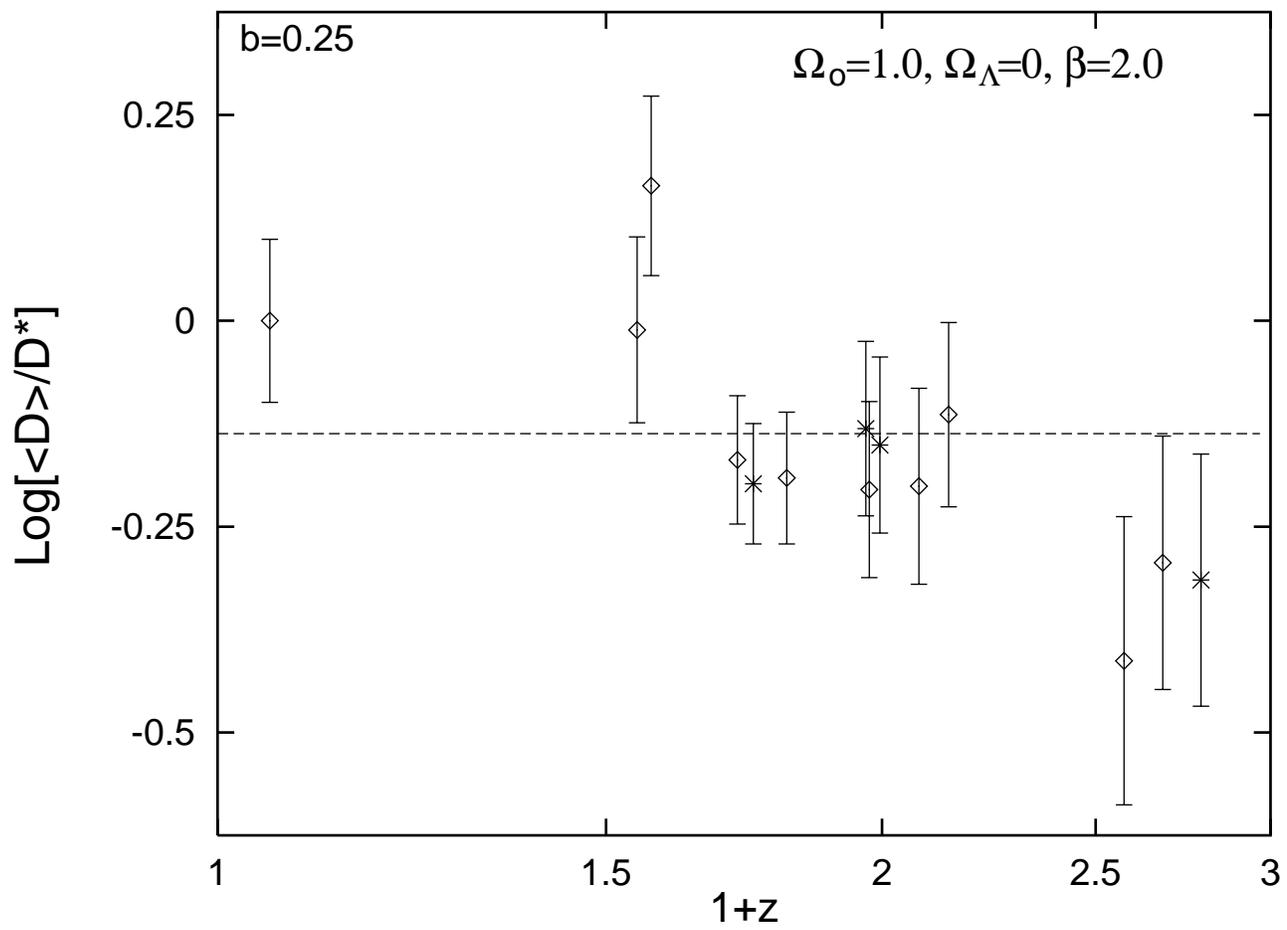
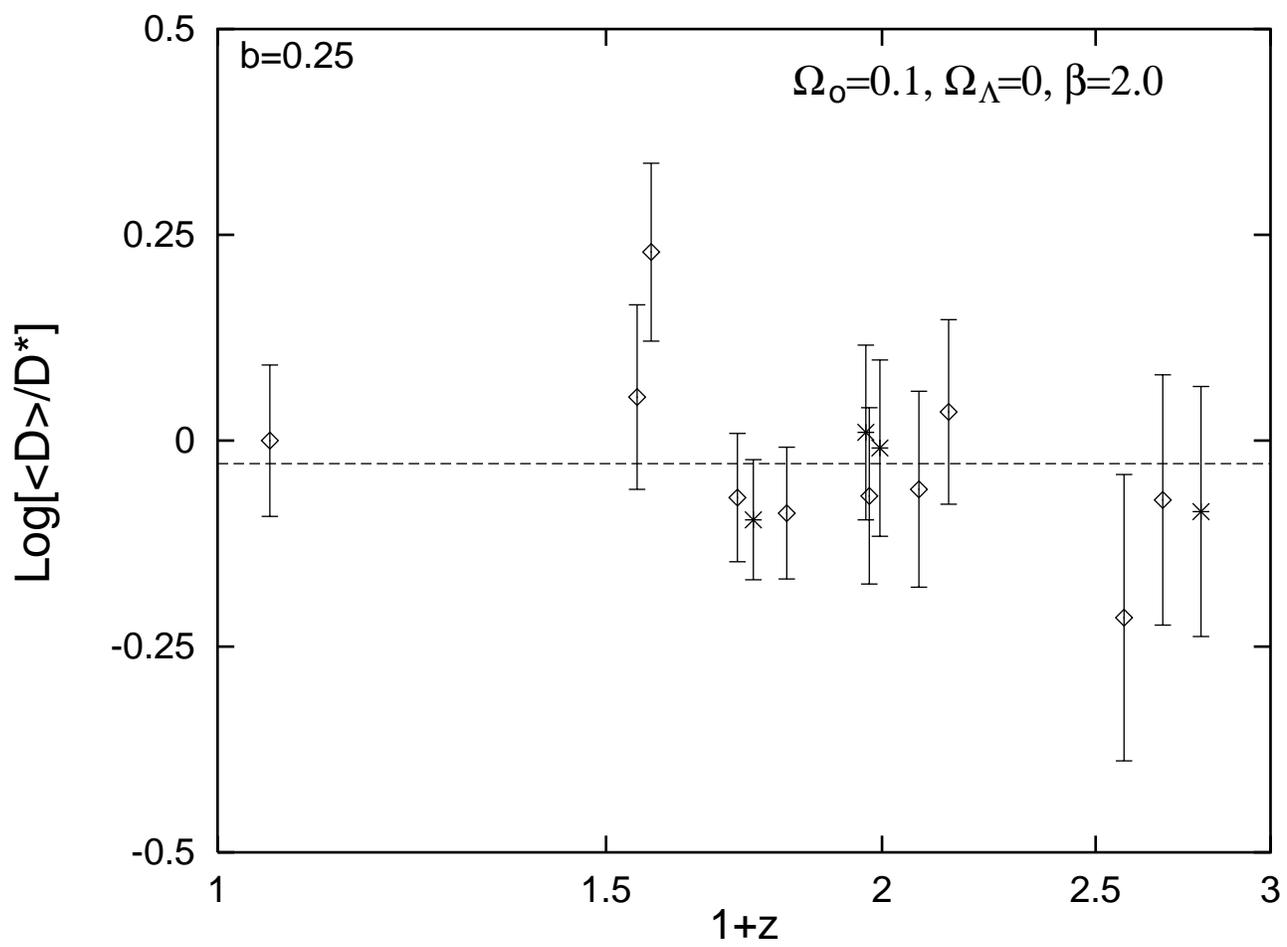
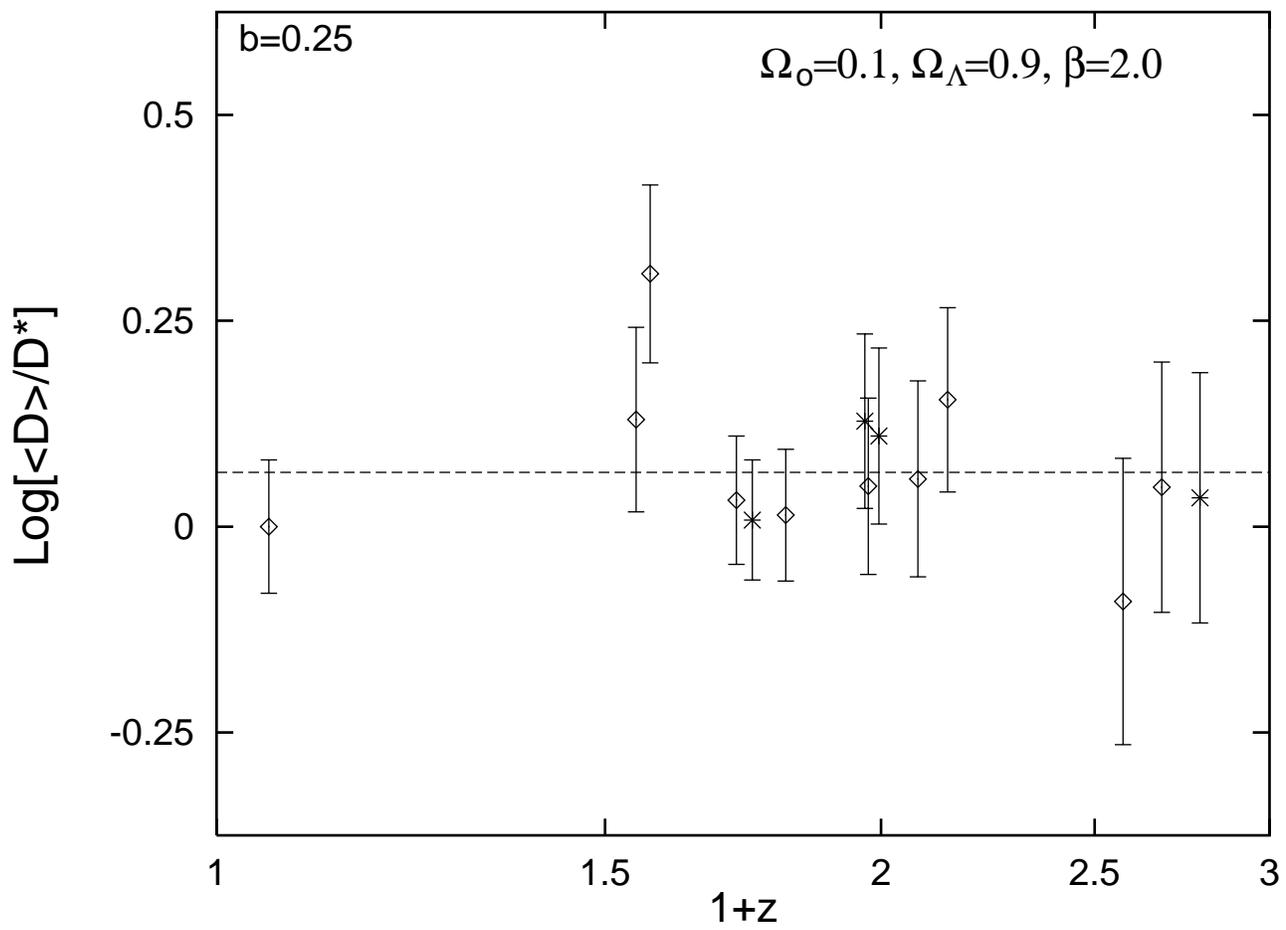
\end{document}